\documentclass[aps,pra,twocolumn,amssymb,amsfonts,floatfix,showpacs,]{revtex4}
\usepackage{bm}
\usepackage{dcolumn}
\usepackage{bm}
\usepackage{mathbbol}
\usepackage{graphicx}
\usepackage{epstopdf} 
\usepackage{amsmath,amsthm}
\usepackage{times}
\usepackage{color}
\usepackage{lipsum}

\begin{document}
\title{Excited-state quantum phase transitions in many-body systems with infinite-range interaction: Localization, dynamics, and bifurcation}

\author{Lea F. Santos}
\affiliation{Department of Physics, Yeshiva University, New York, New York 10016, USA}
\author{Marco T\'avora}
\affiliation{Department of Physics, Yeshiva University, New York, New York 10016, USA}
\author{Francisco P\'erez-Bernal}
\affiliation{Grupo de investigaci\'on en F\'{\i}sica Molecular, At\'omica y Nuclear (GIFMAN-UHU), Unidad Asociada al CSIC. Depto.\ de Ciencias Integradas, Universidad de Huelva, 21071 Huelva, SPAIN}

\date{\today}

\begin{abstract}
Excited state quantum phase transitions (ESQPTs) are generalizations of quantum phase transitions (QPTs) to excited levels. They are associated with local divergences in the density of states. Here, we investigate how the presence of an ESQPT can be detected from the analysis of the structure of the Hamiltonian matrix, the level of localization of the eigenstates, the onset of bifurcation, and the speed of the system evolution. Our findings are illustrated for a Hamiltonian with infinite-range Ising interaction in a transverse field. This is a version of the Lipkin-Meshkov-Glick (LMG) model and the limiting case of the one-dimensional spin-1/2 system with tunable interactions realized with ion traps.  From our studies for the dynamics, we uncover similarities between the LMG and the noninteracting XX models. 
\end{abstract}


\maketitle


\section{Introduction}
Quantum phase transitions (QPTs) correspond to abrupt changes in the character of the ground state of a system when a control parameter reaches a critical point~\cite{CarrBook,SachdevBook}. Strictly, they occur in the thermodynamic limit, but scaling analysis of finite systems can indicate their presence. The nature of the QPTs is determined according to Ehrenfest's classification of thermodynamic phase transitions as transitions of first order, second order, and so on~\cite{Gilmore1979,Feng1981,Jaeger1998}. QPTs have received significant attention by recent experiments with ultracold gases~\cite{Greiner2002,Bloch2008,Baumann2010}.

Excited state quantum phase transition (ESQPTs) refer to QPTs that take place at the excited levels~\cite{Cejnar2006,Caprio2008}. In systems that exhibit an ESQPT, the vanishing gap between the ground state and the first excited state, characteristic of ground state QPTs, does not occur in isolation, but in conjunction with the clustering of the levels near the ground state. This local divergence of the density of states propagates to higher excitation energies as the control parameter increases beyond the ground state critical point. 

ESQPTs have been verified in various models, including molecular vibron~\cite{Caprio2008,Bernal2008}, nuclear interacting boson~\cite{Cejnar2009}, Jaynes-Cummings~\cite{Fernandez2011,Fernandez2011b}, kicked-top \cite{Bastidas2014}, Lipkin-Meshkov-Glick (LMG)~\cite{Cejnar2009,Fernandez2009,Yuan2012}, and Dicke~\cite{Fernandez2011,Fernandez2011b,Brandes2013} models. In the last two cases, the density of states was found analytically~\cite{Ribeiro2008,Brandes2013}.  ESQPTs are not exclusive to integrable models; precursors of the transition persist even in the chaotic domain~\cite{Fernandez2011b,Bastarrachea2014b,ChavezCarlosARXIV,Stransky2015}.
Experimental signatures of ESQPTs were found in the bending motion of different molecular species ~\cite{Winnewisser2005, Zobov2006,Larese2011, Larese2013}, superconducting microwave billiards~\cite{Dietz2013}, and spinor condensates~\cite{Zhao2014}.

Few works exist about the effects of ESQPTs on the system dynamics~\cite{Relano2008,Fernandez2009,Fernandez2011,Engelhardt2015,Puebla2015}. In Refs.~\cite{SantosBernal2015,Bernal2016}, we showed that the time evolution of an initial state with energy close to the ESQPT critical point can be exceedingly slow. These results are general and valid for any Hamiltonian with a ${\rm U}(n+1)$ algebraic structure that has limiting ${\rm SO}(n+1)$ and ${\rm U}(n)$ dynamical symmetries,
\begin{equation}
H_{{\rm U}(n+1)} = (1-\xi) H_{{\rm U}(n)}  + \frac{\xi}{N} H_{{\rm SO}(n+1)} ,
\end{equation}
where $\xi$ is the control parameter and $N$ is the system size.
The ${\rm U}(n+1)$ Hamiltonian in the bosonic form with $n\geq1$ represents the one-dimensional [${\rm U}(2)$], two-dimensional [${\rm U}(3)$], and three-dimensional  [${\rm U}(4)$] limits of the vibron model~\cite{Iachello1981,IachelloBook,Iachello1996,Bernal2005,Bernal2008}. These models are used to describe the vibrational spectra of molecules. The ${\rm U}(2)$ Hamiltonian corresponds to the LMG model~\cite{Lipkin1965a,Lipkin1965b,Lipkin1965c}, introduced in nuclear physics, and since then used in various contexts, from studies of Bose-Einstein condensates to entanglement.

In this work, we focus on the LMG model and extend the results of Refs.~\cite{SantosBernal2015,Bernal2016}. We concentrate on the spin version of the model. It corresponds to an infinite-range Ising interaction [${\rm SO}(2)$ part of the Hamiltonian] in a transverse field [${\rm U}(1)$ part of the Hamiltonian]. This limit of all to all coupling is nearly reached with experiments with ion traps~\cite{Jurcevic2014,Richerme2014}, where the range of the interaction can be tuned. These experiments study the dynamics of the spin system for the same initial states that we consider here. 

We show that at the ESQPT critical point, the eigenstates of the LMG model are highly localized in the ground state of the ${\rm U}(1)$ part of the Hamiltonian. As a consequence, the evolution of this particular basis vector under the LMG Hamiltonian is very slow. The presence of the ESQPT can therefore be detected by analyzing the structure of the eigenstates and the speed of the evolution of ${\rm U}(1)$ basis vectors. The second alternative could be tested with the above mentioned experiments with ion traps~\cite{Jurcevic2014,Richerme2014}. 

A third alternative to identify the presence of the ESQPT that we explore here is the bifurcation phenomenon. It refers to the sudden change in the value of the total magnetization in the direction of the Ising interaction, which occurs at the critical point. Below the energy of the ESQPT, the eigenstates are degenerate, each having a positive or negative value of the total magnetization. Above the critical point, the magnetization of all states becomes zero. Bifurcations similar to this one have been studied experimentally as a function of the control parameter~\cite{Zibold2010,Ferreira2013,Trenkwalder2016}. Here, we analyze how the bifurcation emerges as a function of the excitation energies, while the control parameter is kept fixed and above the QPT critical point. 

Our studies of the dynamics of the LMG model reveals similarities between this model, which has infinite-range interaction, and the XX model with a single excitation, which has only nearest-neighbor couplings. Specifically, the energy distributions of several initial states corresponding to ${\rm U}(1)$ basis vectors are analogous for both systems, which results in equivalent time evolutions. Relationships between the LMG and other integrable models have been explored before~\cite{Amico2008,Lerma2013}, specially in the context of scaling behaviors of the entanglement entropy~\cite{Latorre2005,Barthel2006}. 

The analogy with the XX model motivated a closer look at the structure of the Hamiltonian matrix of the LMG model. From this study, we show that the ESQPT critical energy can be identified even before diagonalization, by simply comparing the spacings between neighboring energy levels and their coupling strengths.

The work is divided as follows. Section II describes the LMG model and gives the Hamiltonian elements in the ${\rm U}(1)$ and in the ${\rm SO}(2)$ bases. Section III provides results for the eigenvalues, structures of the eigenstates, and the magnetizations. It is at this point that we discuss the onset of localized states and bifurcation. Section IV investigates the dynamics under the LMG Hamiltonian for different initial states, establishes a connection between the LMG and XX models, and analyzes the structure of the LMG Hamiltonian matrix. Details about the XX model are found in Appendix A. Final remarks are presented in Sec.~V.

\section{Model}

One-dimensional lattices of interacting spins-1/2 described by the following Hamiltonian,
\begin{equation}
H_{\text{s}}^{(\alpha)} =  B \sum_{i=1}^N \sigma_i^z 
+  \sum_{i<j} \frac{J}{|i-j|^{\alpha}} \sigma_i^x  \sigma_j^x  ,
\label{Hlmg}
\end{equation}
have been recently realized with trapped ions~\cite{Jurcevic2014,Richerme2014}. Above, $\hbar=1$, $\sigma_i^{x,z}$ are Pauli matrices acting on sites $i$,  $N$ is the total number of sites, $B$ is the amplitude of the external field, and $J$ is the coupling parameter. In the experiments, the range of the interaction, determined by $\alpha$, can be tuned from $\alpha=3$ to $\alpha$ very close to zero. The case of infinite-range interaction, $\alpha=0$, corresponds to a version of the LMG model~\cite{Cejnar2009,Yuan2012}.

Hamiltonian (\ref{Hlmg}) for $\alpha=0$ can be written in the form below~\cite{Cejnar2009,Yuan2012},
\begin{equation}
H_{\text{s}}^{(\alpha=0)} =  (1-\xi) \left( \frac{N}{2} + \sum_{i=1}^N S_i^z \right)  
-  \frac{4\xi}{N}  \sum_{i, j=1}^{N} S_i^x S_j^x  ,
\label{Hlmg2}
\end{equation}
where spin operators $S_i^{x,z}$ are used. The necessary steps to reach Eq.~(\ref{Hlmg2}) are:  multiply both terms in $H_{\text{s}}^{(\alpha)}$ by 2, add the constants $2BN$ and $JN$, and then use a single control parameter $\xi$, so that $4B= (1-\xi)$ and $J=-\xi/N$. Note that to guarantee that $H_{\text{s}}^{(\alpha=0)}$ is intensive, the interaction term is rescaled with $1/N$. 

In general, the Hamiltonian matrix from Eq.~(\ref{Hlmg}) has total dimension $2^N$, but when $\alpha=0$ [Eq.~(\ref{Hlmg2})], its effective size reduces to  $N+1$. All $N!/(N_{up}! N_{down}!)$ states with $N_{up}$ spins pointing up in the $z$-direction and $N_{down}$ spins pointing down become degenerate. The Hamiltonian can then be written in terms of the total spin in the $z$-direction, ${\cal S}_z = \sum_{i=1}^{N} S_i^z$, and  the total spin in the $x$-direction, ${\cal S}_x = \sum_{i=1}^{N} S_i^x$, as
\begin{equation}
H_{\text{s}} = (1-\xi) \left( \frac{N}{2} + {\cal S}_z \right) -  \frac{4\xi}{N} {\cal S}_x^2 .
\label{HtotalS}
\end{equation}

The LMG Hamiltonian $H_{\text{s}}$ (\ref{HtotalS}) has a ${\rm U}(2)$ algebraic structure with two limiting dynamical symmetries represented by the ${\rm U}(1)$ subalgebra, when $\xi=0$, and the ${\rm SO}(2)$ subalgebra, when $\xi=1$. The eigenstates of the ${\rm U}(1)$ part of the Hamiltonian correspond to the states $|s \, m_z\rangle$ and those of the ${\rm SO}(2)$ part are the states $|s \, m_x\rangle$, where $s=N/2$ is  the total spin quantum number and $m_{z(x)}$ is the total magnetization in the $z$($x$)-direction, with $-N/2 \leq m_{z(x)} \leq N/2$.

The elements of the Hamiltonian matrix in the ${\rm U}(1)$ basis are given by
\begin{eqnarray}
&&\langle s \, m_z | H_{\text{s}} | s \, m_z \rangle =  \left( \frac{N}{2}+m_z \right)  \left( 1 -  2\xi + 2 \frac{\xi m_z}{N} \right) -\xi , \nonumber \\
&&\langle s \, m_z +2| H_{\text{s}} | s \, m_z  \rangle = -\frac{\xi}{N} \sqrt{\left(\frac{N}{2}+m_z+2\right) }   \nonumber \\
&&\times \sqrt{ \left( \frac{N}{2}+m_z+1 \right) \left(\frac{N}{2} -m_z \right) \left(\frac{N}{2} -m_z -1\right)} . \nonumber
\end{eqnarray}
$H_{\text{s}}$~(\ref{HtotalS}) conserves parity, $(-)^{s+m_z}$ \cite{Cejnar2009}, so the matrix is split in two blocks, one of dimension $D_{even}=N/2+1$ with even parity and the other of dimension $D_{odd}=N/2$ and odd parity. 

In the ${\rm SO}(2)$ basis, the elements of the Hamiltonian matrix are
\begin{eqnarray}
&&\langle s \, m_x | H_{\text{s}} | s \, m_x \rangle =  -\frac{4\xi}{N} m_x^2 + (1-\xi) \frac{N}{2} , \nonumber \\
&&\langle s \, m_x +1| H_{\text{s}} | s \, m_x  \rangle = \frac{\xi-1}{2} \sqrt{\left(\frac{N}{2}-m_x\right) \left(\frac{N}{2}+m_x+1\right) }   . \nonumber
\end{eqnarray}

Hamiltonian~(\ref{HtotalS}) may also be written in a bosonic form. The Holstein-Primakoff mapping is not suitable here, because the total number of bosons in this representation is not conserved. Instead, the Schwinger representation is more appropriate,
\begin{eqnarray}
&& {\cal S}_z = \sum_{i=1}^N S_i^z =t^{\dagger} t - \frac{N}{2} = n_t - \frac{N}{2}  \\
&& {\cal S}^{+}  =\sum_{i=1}^N S_i^+ = t^{\dagger} s = ({\cal S}^-)^{\dagger}.
\end{eqnarray}
The resulting Hamiltonian describes a system with two species of scalar bosons,  boson $s$ and boson $t$,
\begin{equation}
H_{\text{b}} =  (1-\xi) t^{\dagger} t - \frac{\xi}{N} ( t^{\dagger} s + s^{\dagger} t )^2 ,
\label{Hbosons}
\end{equation}
where $N$ is the conserved total number of bosons $N=n_t + n_s$. The elements of the Hamiltonian matrix in the basis
\begin{equation}
|N n_t\rangle = \frac{(t^{\dagger})^{n_t} (s^{\dagger})^{N-n_t} } {\sqrt{ n_t! (N-n_t)!} } |0 \rangle ,
\end{equation}
where $0\leq n_t \leq N$ and $|0\rangle$ is the vacuum state~\cite{Fernandez2009,Yuan2012,FrankBook,IachelloBook2},
are analogous to those for the $|s \, m_z\rangle$ basis, substituting $m_z$  with  $n_t-N/2$.

The LMG Hamiltonian shows a second-order QPT at the critical point $\xi_c=1/5$. The ESQPT occurs for $\xi >\xi_c$.

\section{Eigenvalues, eigenstates, and observables}
\begin{figure}[htb]
\centering
\includegraphics*[width=3.5in]{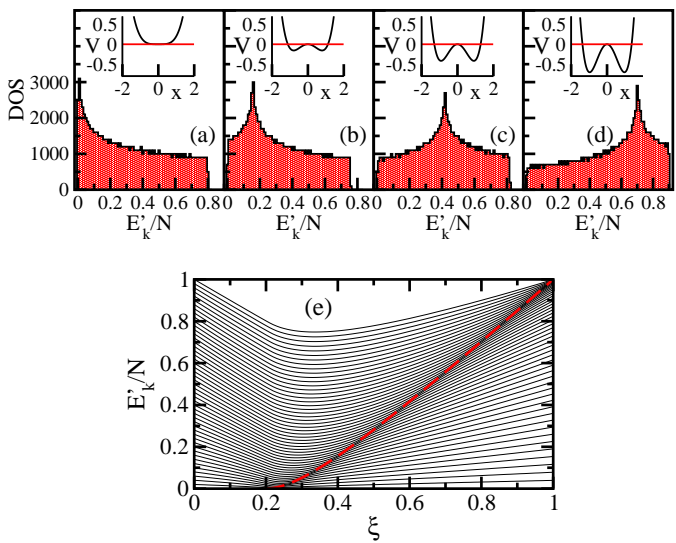}
\caption{(Color online) Top panels: Normalized density of states for $H_s$ (\ref{HtotalS}) with $\xi_c=0.2$ (a), $\xi=0.4$ (b), $\xi=0.6$ (c) and $\xi=0.8$ (d), $N=2000$. The corresponding classical potentials [Eq.~(\ref{eq:potential})] are shown in the insets. Bottom panel (e): Normalized excitation energies \textit{vs} $\xi$, $N = 100$.  The separatrix [Eq.~(\ref{eq:separatrix})] is indicated with the dashed line. All panels: even parity sector. Arbitrary units.}
\label{fig:DOS}
\end{figure}

ESQPTs are characterized by the clustering of the eigenvalues around the energy $E_{\text{ESQPT}}$ of the critical point. This is illustrated with the density of states for the LMG model in Figs.~\ref{fig:DOS} (a), (b), (c), and (d) for $\xi=0.2, 0.4, 0.6, 0.8$, respectively. There, and throughout this paper, we subtract from the eigenvalues $E_k$ the energy of the ground state $E_0$ and deal with  $E_k'= E_k-E_0$. From those four panels, one sees that the peak of the distribution moves to higher energies as $\xi$ increases from the QPT critical point ($\xi_c=0.2$) up. The value of $E_{\text{ESQPT}}$ therefore depends on $\xi$.

\subsection{Separatrix and semiclassical approximation}
The dependence of the value of $E_{\text{ESQPT}}$  on the control parameter is visible also in Fig.~\ref{fig:DOS} (e), where we plot the normalized excitation energies $E_k'/N$ for all levels versus $\xi$.
The dashed line in that panel follows the clustering of the eigenvalues. This line corresponds to the separatrix that marks the ESQPT. Its equation,
\begin{equation}
E_{\text{ESQPT}}  (\xi)= \frac{(1-5\,\xi)^2}{16\,\xi},
\label{eq:separatrix}
\end{equation}
is obtained in the mean-field approximation (limit of very large $N$), as summarized below~\cite{Bernal2008,Caprio2008}. 

Using Glauber coherent states, we can write the classical limit of Hamiltonian \eqref{Hbosons}  in terms of coordinate and momenta as (\cite{Caprio2008} and references therein),
\begin{equation}
H_{class} = \frac{1-\xi}{2N^2} p^2 + \frac{\xi}{N^2} x^2 p^2 + V(x), 
\end{equation}
where the potential is 
\begin{equation}
V(x) = \frac{1-5\xi}{2}x^2 + \xi x^4 . 
\label{eq:potential}
\end{equation}
We can also use projective coherent states~\cite{IachelloBook2,IachelloBook3} and put momenta equal to zero to obtain the classical energy functional associated with Hamiltonian \eqref{Hbosons}, which is given by  (\cite{Bernal2008} and references therein),
\begin{equation}
{\cal E}_\xi(x) = (1-\xi)\frac{x^2}{1+x^2} -\xi \frac{4x^2}{(1+x^2)^2}~.
\label{eq:functional}
\end{equation}
Either from Eq.~(\ref{eq:potential}) or from Eq.~(\ref{eq:functional}), 
we see that when $\xi \leq \xi_c=1/5$, the potential has a minimum at $x=0$, which is quadratic for $\xi <\xi_c$ and quartic for $\xi =\xi_c$ [inset of Fig.~\ref{fig:DOS} (a)]. For $\xi >\xi_c$, the potential has a double-well shape [insets of Figs.~\ref{fig:DOS} (b), (c), and (d)], with minima at $x= \pm \sqrt{(5 \xi -1)/(4\xi)}$, while $x=0$ is now a maximum. The energy difference between the maximum value $V(x=0)$ and the minimum value $V(x=\pm \sqrt{(5 \xi -1)/(4\xi)})$ marks the ESQPT critical energy and leads to the equation of the separatrix [Eq.~(\ref{eq:separatrix})]. At excitation energies equal to $E_{\text{ESQPT}}(\xi) = V(x=0) - V \left(x = \pm \sqrt{(5 \xi -1)/(4\xi)}\right) = (1-5\xi)^2/(16\xi)$,  the origin, which was prohibited for  $E<E_{\text{ESQPT}}$ due to the potential barrier, can now be reached. 

The emergence of ESQPTs can therefore be understood from the double-well potential.  For energies very close to the top of the potential barrier, the classical velocity becomes very small, indicating that a system with energy $\sim E_{\text{ESQPT}}$ spends a long time in the vicinity of $x=0$. The appearance of such stationary point is associated with the singularity in the density of states marked by the separatrix~\cite{Cejnar2006,Stransky2014,Child1998}.

The above classical picture helps the understanding of the structure of the eigenstates of the algebraic quantum model.  The ${\rm U}(1)$-part of the Hamiltonian corresponds to a truncated one-dimensional harmonic oscillator, where the ground state  $n_t=0$ ($m_z=-N/2$) has a large probability to be found at the origin. Since, in analogy with the above discussion, the 
eigenstates with energies very close to $E_{\text{ESQPT}}$ are also likely to be found around $x=0$,  they must be highly localized in the ${\rm U}(1)$-ground state. This is corroborated by our results for the eigenstates in the next subsection. 

\subsection{Structure of the Eigenstates in the ${\rm U}(1)$ basis}
Written in the ${\rm U}(1)$ basis, the eigenstates with energies below the separatrix, $E_k'/N<E_{\text{ESQPT}}$, have a structure closer to that of the eigenstates of the ${\rm SO}(2)$-Hamiltonian, while those with energies above the separatrix are more similar to the eigenstates of the ${\rm U}(1)$-Hamiltonian~\cite{Caprio2008}. The eigenstates with energy very close to the separatrix, $E_k'/N \sim E_{\text{ESQPT}}$,  are the ones at the point of transition from one dynamical symmetry to the other and they are highly localized in the ${\rm U}(1)$-ground state, which has $m_z=-N/2$ ($n_t=0 $). 

\subsubsection{Components of the Eigenstates in the ${\rm U}(1)$ basis}

\begin{figure}[htb]
\centering
\includegraphics*[width=3.5in]{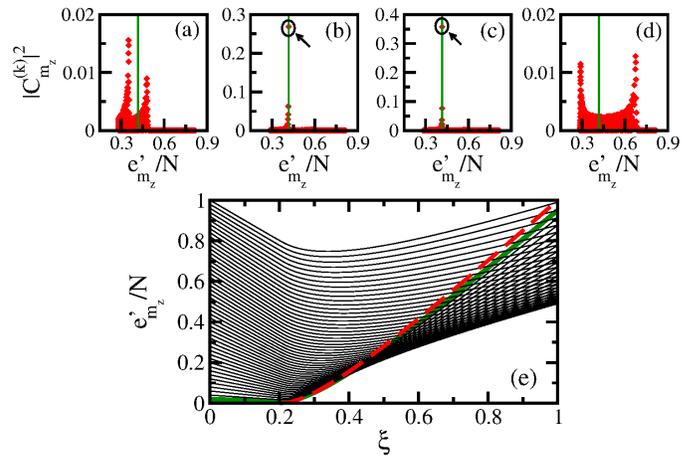}
\caption{(Color online) Top panels: squared coefficients $|C_{m_z}^{(k)}|^2$ of the eigenstates $|\psi_k\rangle$ written in the ${\rm U}(1)$ basis  {\em vs} the energies of the corresponding basis vectors; $\xi =0.6$, $N = 2000$. The eigenstates chosen have energies $E_k'/N = 0.2515$ (a), $0.4163$ (b) [second closest to the ESQPT critical point], $0.4166$ (c) [closest one to the ESQPT critical point], and $0.5764$ (d).  Vertical lines indicate the separatrix, $E_{\text{ESQPT}}=0.4167$. Bottom panel (e): Normalized energy of the ${\rm U}(1)$ basis vectors in the total Hamiltonian \textit{vs} $\xi$, $N = 100$.  The separatrix [Eq.~(\ref{eq:separatrix})] is indicated with the dashed line and the energy of the ${\rm U}(1)$-ground state $e_{-N/2}'/N$ with the thick solid line. All panels: even parity sector.  Arbitrary units.}
\label{fig:PSI}
\end{figure}

In Figs.~\ref{fig:PSI} (a), (b), (c), and (d), we show the structures of four eigenstates written in the ${\rm U}(1)$ basis. $|C_{m_z}^{(k)}|^2$ is the probability to find the eigenstate $|\psi_{k}\rangle = \sum_{m_z=-N/2}^{N/2} C_{m_z}^{(k)} |s \, m_z \rangle$ in the basis vector $|s \, m_z\rangle$ and $e_{m_z}' = \langle s \, m_z | H_{\text{s}} | s \, m_z \rangle - E_0$ is the energy of the basis vector in the total Hamiltonian shifted by the ground state energy of $H_{\text{s}}$. The energy of the eigenstate in Fig.~\ref{fig:PSI} (a) [(d)] is below [above] the separatrix; there are several basis vectors contributing to this eigenstate and they mostly have low [high] energies. In Figs.~\ref{fig:PSI} (b) and (c) we show, respectively, the eigenstate with the second closest and the closest normalized energy to $E_{\text{ESQPT}}$. These states are highly localized in the ${\rm U}(1)$-ground state ($m_z=-N/2$). The point for $|C_{-N/2}^{(k)}|^2$ is indicated with an arrow in the figures. Compare also the $y$-axis scales in Figs.~\ref{fig:PSI} (b) and (c) with Figs.~\ref{fig:PSI} (a) and (d).

The localization of the eigenstates with $E'_k/N \sim E_{\text{ESQPT}}$ in the ${\rm U}(1)$-ground state can be anticipated by computing the energy $e_{-N/2}'/N$, which is also very close to $E_{\text{ESQPT}}$. As shown in Fig.~\ref{fig:PSI} (e), $e_{-N/2}'/N$ follows the separatrix as $\xi$ increases. Note that for a given $N$, the difference $E_{\text{ESQPT}}-e_{N/2}'/N$ increases with $\xi$, but at the same, for a fixed $\xi>\xi_c$, this difference decreases with $N$. 

For $\xi<\xi_c$, $ | s \, m_z\!\!=\!\!-N/2 \rangle$ is the basis vector with the lowest energy. As $\xi$ increases above $\xi_c$, this state is carried up in energy and $e_{-N/2}'/N$ gets above the energy of some of the basis vectors with $m_z>-N/2$. The number of states with $e_{m_z}'<e_{-N/2}'$ increases with $\xi$. At $\xi=1$, the energies of all ${\rm U}(1)$ basis vectors are below $e_{-N/2}'$, apart from $m_z=N/2$, which becomes degenerate with it.

\begin{figure}[htb]
\centering
\includegraphics*[width=3.5in]{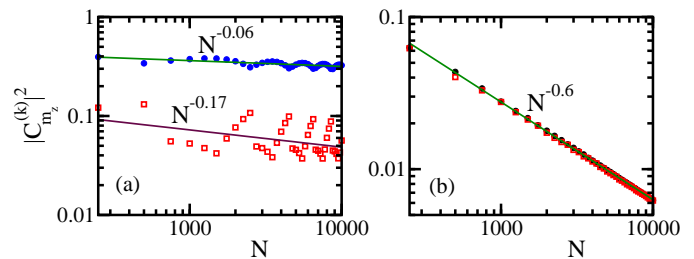}
\caption{(Color online) Log-log plots of the  largest components (filled circles) and the second largest components (empty squares) $|C_{m_z}^{(k)}|^2$ {\em vs} $N$ for the eigenstate that is most localized in the ${\rm U}(1)$-ground state [its energy $E_{loc}'/N$ is very close to the $E_{\text{ESQPT}}$] (a) and for the eigenstate at a position $D_{even}/4$ above $E_{loc}'/N$ (b). Solid lines are fittings with indicated powerlaw decays, $\xi=0.6$ and even parity.}
\label{fig:Ck}
\end{figure}

In Figs.~\ref{fig:Ck} (a) and (b), we study the dependence of the largest and the second largest  components $|C_{m_z}^{(k)}|^2$ on the system size for two different eigenstates. In Fig.~\ref{fig:Ck} (a), we select the eigenstate that is most localized in $ | s \, m_z\!\!=\!\!-N/2 \rangle$. The energy of this eigenstate is very close to $E_{\text{ESQPT}}$, although for some system sizes, it is not the closest one to the separatrix.  The figure shows that the largest component decays slower with $N$ than the second largest one, indicating that $ | s \, m_z\!\!=\!\!-N/2 \rangle$ is indeed the preferred basis vector for any system size. In Fig.~\ref{fig:Ck} (b), we choose an eigenstate with energy above the separatrix. In this case, the magnitudes of the largest and second largest components practically coincide, indicating no preference for a particular basis vector. These components decrease much faster with $N$ than those two for the localized state.

\subsubsection{Level of Localization of the Eigenstates in the ${\rm U}(1)$ basis}

The change in the structure of the eigenstates written in the ${\rm U}(1)$ basis as they approach the separatrix signals the existence of an ESQPT. To evaluate this change, we may use quantities, such as the participation ratio (PR) or the Shannon (information) entropy~\cite{ZelevinskyRep1996,Kota2001,Santos2005loc,Gubin2012}, that measure the level of localization of the eigenstates in a chosen basis. The $P$ is defined as
\begin{equation}
\text{P}^{(k)}_{{\rm U}(1)} =\frac{1}{\sum_{m_z} | C_{m_z}^{(k)}|^4}.
\label{eq:IPR}
\end{equation}
A large value  indicates an extended state in the chosen basis and a small value, a localized state. When $\xi=0$, the eigenstates coincide with the ${\rm U}(1)$-basis vectors, so $\text{P}^{(k)}_{{\rm U}(1)} =1$.

\begin{figure}[htb]
\centering
\includegraphics*[width=3.3in]{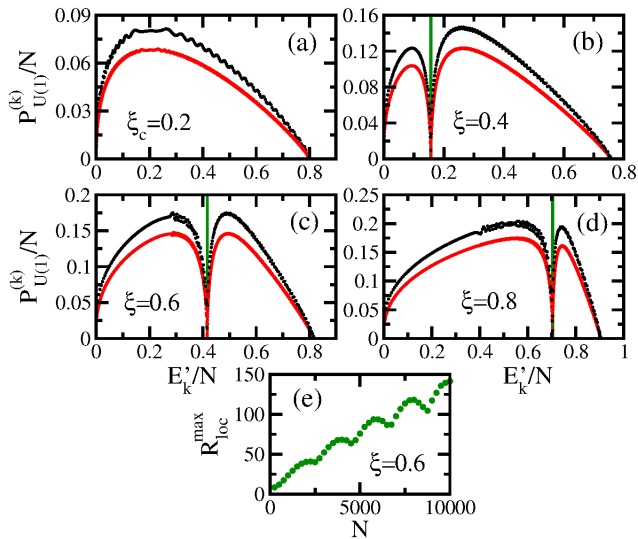}
\caption{(Color online) Panels (a), (b), (c), and (d): Participation ratio of all the eigenstates of the even parity sector written in the ${\rm U}(1)$ basis; $N=500$ (dark curve) and $2000$ (light curve). Vertical lines mark the $E_{\text{ESQPT}}$ obtained from Eq.~(\ref{eq:separatrix}). Panel (e): dependence on $N$ of the ratio $R^{\text{max}}_{\text{loc}} =\text{P}^{\text{max}}_{{\rm U}(1)}/ \text{P}^{\text{loc}}_{{\rm U}(1)}$ between the participation ratio of the most delocalized state, $\text{P}^{\text{max}}_{{\rm U}(1)}$,  and the P of the most localized state in the ${\rm U}(1)$ ground state, $\text{P}^{\text{loc}}_{{\rm U}(1)}$. Arbitrary units.}
\label{fig:PR}
\end{figure}

In Figs.~\ref{fig:PR} (a)-(d), we show $\text{P}^{(k)}_{{\rm U}(1)}/N$ for all eigenstates~\cite{footPR}. Each panel has a different value of the control parameter. For $0\leq \xi \leq \xi_c$, P is a smooth function of energy, indicating more localized states at the edges, as seen in Fig.~\ref{fig:PR} (a). Above the critical point [Figs.~\ref{fig:PR} (b), (c), and (d)], the eigenstates remain localized at the edges of the spectrum, but the same happens also for the states with energies close to $E_{\text{ESQPT}}$. This causes the dip in the value of $\text{P}^{(k)}_{{\rm U}(1)}$ for $E_k'/N\sim E_{\text{ESQPT}}$, as seen in the figures. The P serves therefore as an order parameter for ESQPTs.

Overall, $\text{P}^{(k)}_{{\rm U}(1)}/N$ decreases with system size for all eigenstates, indicating that they are far from being ergodic. Ergodicity implies that $\text{P}^{(k)}_{{\rm U}(1)} \propto N$. However, the participation ratio of the most localized state in the ${\rm U}(1)$ ground state, $\text{P}^{\text{loc}}_{{\rm U}(1)}$, decays faster with $N$ than the P of the most delocalized state, $\text{P}^{\text{max}}_{{\rm U}(1)}$. This is clearly seen in Fig.~\ref{fig:PR} (e), which shows the dependence on the system size of the ratio $R^{\text{max}}_{\text{loc}} =\text{P}^{\text{max}}_{{\rm U}(1)}/ \text{P}^{\text{loc}}_{{\rm U}(1)}$. Thus, the level of localization of the eigenstates with energies very close to the separatrix gets more pronounced with $N$ than for other generic eigenstates.

\subsection{Structure of the Eigenstates in the ${\rm SO}(2)$ basis}

\begin{figure}[htb]
\centering
\includegraphics*[width=2.5in]{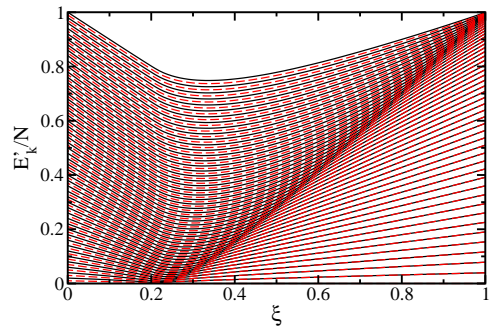}
\caption{(Color online) Normalized excitation energies {\em vs} $\xi$ for all $N+1$ eigenstates, including both parities, one parity is indicated with solid lines and the other with dashed lines; $N=100$.}
\label{fig:parities}
\end{figure}

An important aspect of the eigenstates below the separatrix is that those with the same value of $|m_x|$ are degenerate. This can be explained as follows. The ${\rm SO}(2)$-part of $H_s$ is given by the square of the operator ${\cal S}_x$; the eigenstates of ${\cal S}_x^2$ with the same value of $|m_x|$ are degenerate. The same occurs to the eigenstates of $H_s$ that have energy below the separatrix, since they are closer to the ${\rm SO}(2)$-symmetry. In contrast, above the separatrix, where the eigenstates of $H_s$ are closer to the ${\rm U}(1)$-symmetry, the degeneracy is lifted. In this region $m_x=0$. [This sudden change in the value of $m_x$ at the separatrix is related to the bifurcation phenomenon that is described in the next subsection.] In Fig.~\ref{fig:parities},  we consider all $N+1$ eigenvalues of the Hamiltonian $H_s$ (\ref{HtotalS}). The separatrix clearly marks the point where pairs of eigenstates with different parity are distinguished by energy (above the separatrix), from those that are degenerate (below the separatrix).  

\begin{figure}[htb]
\centering
\includegraphics*[width=3.3in]{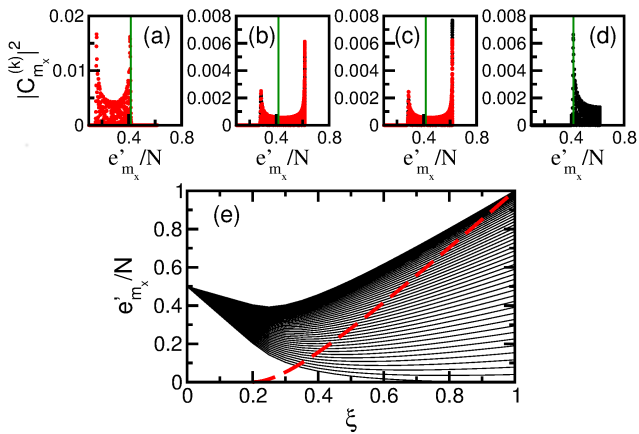}
\caption{(Color online) Top panels: squared coefficients $|C_{m_x}^{(k)}|^2$ of the eigenstates $|\psi_k\rangle$ written in the ${\rm SO}(2)$ basis {\em vs} the energies of the corresponding basis vectors; $\xi =0.6$, $N = 2000$. The eigenstates shown have the energies considered in Fig.~\ref{fig:PSI}: two degenerate states with $E_k'/N = 0.2515$ (a), two with $E_k'/N \sim 0.4163$ (b) [second closest energy to $E_{\text{ESQPT}}$], two with $E_k'/N \sim 0.4166$ (c) [closest energy to $E_{\text{ESQPT}}$], and one state with $E_k'/N = 0.5764$ (d).  Vertical lines mark the ESQPT energy $E_k'/N=0.4167$. Bottom panel (e): Normalized energy of the ${\rm SO}(2)$ basis vectors in the total Hamiltonian \textit{vs} $\xi$, $N = 100$.  The separatrix [Eq.~(\ref{eq:separatrix})] is indicated with the dashed line. Both parities are included. Arbitrary units.}
\label{fig:PSI_SO2}
\end{figure}

In the top panels of Fig.~\ref{fig:PSI_SO2}, we show the structures of the eigenstates with the same energies considered in Fig.~\ref{fig:PSI}, but now written in the ${\rm SO}(2)$ basis. 
$|C_{m_x}^{(k)}|^2$ is the probability to find the eigenstate $|\psi_{k}\rangle = \sum_{m_x=-N/2}^{N/2} C_{m_x}^{(k)} |s \, m_x \rangle$ in the basis vector $|s \, m_x\rangle$ 
 and $e_{m_x}' = \langle s \, m_x | H_{\text{s}} | s \, m_x \rangle - E_0$ is the energy of the ${\rm SO}(2)$ basis vector in the LMG Hamiltonian shifted by the ground state energy of $H_{\text{s}}$. 
In Fig.~\ref{fig:PSI_SO2} (a), the energy is below the separatrix, so there are two degenerate eigenstates perfectly overlapping. They have contributions from basis vectors with energies below $E_{\text{ESQPT}}$. Very close to the separatrix [Figs.~\ref{fig:PSI_SO2} (b) and (c)], the two eigenstates shown in each panel are very similar, but not exactly equal anymore. Above the separatrix [Fig.~\ref{fig:PSI_SO2} (d)], where eigenstates of different parity have different energies, only one eigenstate  is considered, the same one from Fig.~\ref{fig:PSI} (d). In this case, all contributing basis vectors have energy values above the separatrix. 

In the ${\rm SO}(2)$ basis, there is no particularly localized eigenstate, apart from those at the edges of the spectrum. None of the basis vectors has a special role, as the ${\rm U}(1)$-ground state has. In Fig.~\ref{fig:PSI_SO2} (e), we show the energies of all ${\rm SO}(2)$ basis vectors, $e_{m_x}'$, {\em vs} the control parameter. The main effect of increasing $\xi$ is the spreading of the energies of these states. Despite this seemingly lack of special features of the eigenstates in the ${\rm SO}(2)$ basis, the participation ratio can still detect the ESQPT, as discussed next.

\begin{figure}[htb]
\centering
\includegraphics*[width=3.3in]{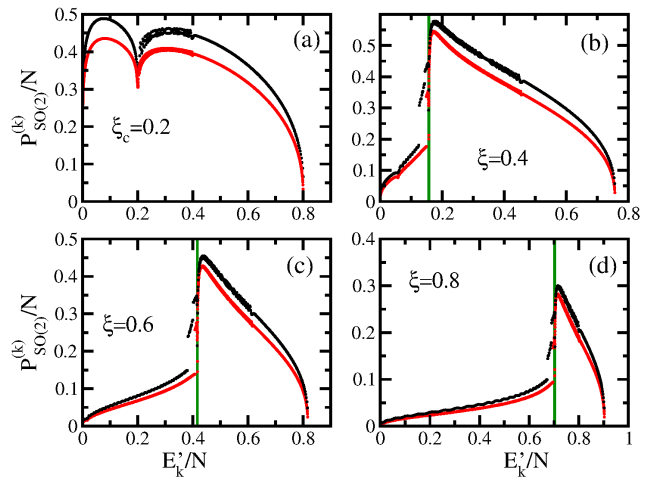}
\caption{(Color online) Participation ratio of all $N+1$ eigenstates of both parity sectors written in the ${\rm SO}(2)$ basis; $N=500$ (dark curve) and $2000$ (light curve). Vertical lines mark the $E_{\text{ESQPT}}$ obtained from Eq.~(\ref{eq:separatrix}). Arbitrary units.}
\label{fig:PR_SO2}
\end{figure}

In Figs.~\ref{fig:PR_SO2} (a), (b), (c), and (d), we show $\text{P}^{(k)}_{{\rm SO}(2)}/N$ for all eigenstates written in the ${\rm SO}(2)$ basis. There is a discontinuity at $E_{\text{ESQPT}}$, above which the eigenstates suddenly become much more delocalized. This is somewhat expected, since the eigenstates above the separatrix are closer to the ${\rm U}(1)$-symmetry than to the ${\rm SO}(2)$-symmetry. The sudden jump to higher values of $\text{P}^{(k)}_{{\rm SO}(2)}/N$, marked by a gap in the values of the participation ratio at the separatrix, may be seen as a signature of the ESQPT. 

\subsection{Observables}

\begin{figure}[htb]
\centering
\includegraphics*[width=3.3in]{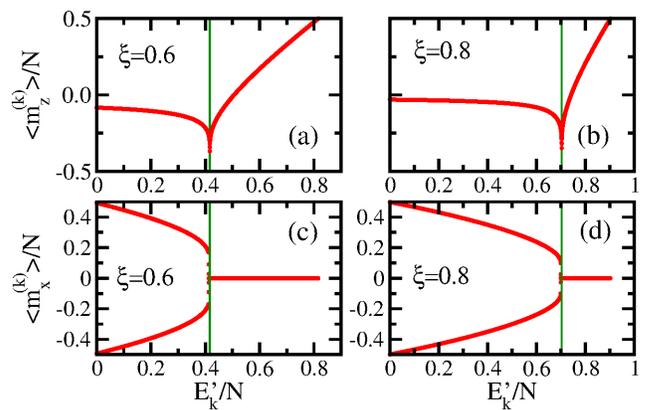}
\caption{(Color online) Top: Normalized total magnetization in the $z$-direction for all eigenstates with even parity. Bottom: Normalized total magnetization in the $x$-direction for all eigenstates of both parities. Vertical lines indicate the separatrix [Eq.~(\ref{eq:separatrix})]; $N=2000$.}
\label{fig:nt}
\end{figure}

A natural consequence of the localization in the ${\rm U}(1)$-ground state of the eigenstates that have energy close to the separatrix is their reduced value of the total magnetization in the $z$-direction.
This is illustrated in Figs.~\ref{fig:nt} (a) and (b), which show the normalized $z$-magnetization, $\langle m_z^{(k)} \rangle/N = \langle \psi_k  | {\cal S}_z | \psi_k \rangle/N$, for all eigenstates. For the states below the separatrix, the range of values of $\langle m_z^{(k)} \rangle/N$ is quite limited and very close to zero. This reflects the proximity of these states to the ${\rm SO}(2)$-symmetry, for which  $\langle s\, m_x  | {\cal S}_z | s\, m_x \rangle/N =0$. At the separatrix, $\langle m_z^{(k)} \rangle/N$ suddenly approaches $-1/2$, which is the value for the ${\rm U}(1)$-ground state. Above the separatrix, a broad range of values are obtained up to $\langle m_z^{(k)} \rangle/N\sim +1/2$. [Similar results were shown for a ${\rm U}(3)$-Hamiltonian in~\cite{Bernal2010}.]

For the normalized total magnetization in the $x$-direction, $\langle m_x^{(k)} \rangle/N = \langle \psi_k  | {\cal S}_x | \psi_k \rangle/N$, a discontinuity also occurs at the separatrix, as seen in Figs.~\ref{fig:nt} (c) and (d). For energies below the separatrix, pairs of degenerate eigenstates have the same magnitude of $|\langle m_x^{(k)} \rangle|/N$. In this energy region, the eigenstates have structures similar to those of the eigenstates of the ${\cal S}_x^2$ operator, that is the ${\rm SO}(2)$-part of the Hamiltonian. Above the separatrix, where the eigenstates are closer to the ${\rm U}(1)$-symmetry, the value of the $x$-magnetization becomes zero. This is an example of the bifurcation phenomenon~\cite{Shchesnovich2009,Diaz2010,Trenkwalder2016,Zibold2010,Ferreira2013}, which has been associated with the presence of QPTs. Figures~\ref{fig:nt} (c) and (d) indicate that it also detects the presence of ESQPTs. The onset of the bifurcation happens not only for the ground state~\cite{Trenkwalder2016} and not only as a function of the control parameter~\cite{Zibold2010,Ferreira2013,Trenkwalder2016}, but also for a fixed $\xi > \xi_c$ as a function of energy.

\section{Quench Dynamics}

From the previous results for the eigenstate expectation values of the magnetizations and the structures of the eigenstates, we may anticipate the dynamics of the LMG model and other systems exhibiting ESQPTs. For instance, due to the localization of the eigenstates with $E'_k \sim E_{\text{ESQPT}}$ in $|s \, m_z\rangle=|s \, -\!\!N/2\rangle$, this basis vector should evolve slowly under $H_s$ (\ref{HtotalS}). We also expect the total $x$-magnetization of an initial state corresponding to $|s \, m_x\rangle=|s \, 0\rangle$ to be dynamically frozen under $H_s$. These predictions, as well as other results, are explored in this section. The main motivation for studying dynamics comes from current experiments with ion traps~\cite{Jurcevic2014,Richerme2014} and optical lattices~\cite{Trotzky2012,Kaden2014}, where dynamics is routinely analyzed.

Here, we study the evolution of different ${\rm U}(1)$ basis vectors and ${\rm SO}(2)$ basis vectors under the LMG Hamiltonian $H_s$ (\ref{HtotalS}) with $\xi$ above the QPT critical point. Having as initial state a ${\rm U}(1)$ basis vector is equivalent to performing an abrupt perturbation (quench), where $\xi$ is initially $0$ and is then suddenly changed to a value $\xi>\xi_c$. Using the ${\rm SO}(2)$ basis vector as initial state corresponds to quenching the control parameter from $\xi=1$ to $\xi>\xi_c$.

The quantities considered for the time evolution analysis are the survival probability and the total magnetizations in the $z$- and $x$-directions. The survival probability of the initial state, also called non-decay probability or fidelity, is given by the absolute square of the overlap between the initial state $|\Psi(0) \rangle=|s \, {\text{ini} }\rangle$  (where $\text{ini}$ stands for a value of $m_z$ or $m_x$) and the evolved state  $|\Psi(t) \rangle$, as
\begin{eqnarray}
F(t) &\equiv& \left| \langle \Psi(0) | \Psi(t) \rangle \right|^2  = \left| \langle \Psi(0) | e^{-i H_s t} | \Psi(0) \rangle \right|^2  
\label{eq:fidelity} \\
&=& \left|\sum_{k} |C_{\text{ini} }^{(k)} |^2 e^{-i E_{k} t}  \right|^2 = 
\left| \int \!\! dE e^{-i E t} \rho_{\text{ini} }(E) \right|^2 .
\nonumber
\end{eqnarray}
Above,  $\rho_{\text{ini} }(E) = \sum_k |C_{\text{ini} }^{(k)} |^2\delta(E-E_k)$ is the energy distribution of $| \Psi(0) \rangle$ weighted by the components  $|C_{\text{ini} }^{(k)} |^2$. One often refers to $\rho_{\text{ini} }(E)$  as strength function~\cite{Borgonovi2016} or local density of states (LDOS); we use the latter term. It is evident from Eq.~(\ref{eq:fidelity}) that the survival probability is the absolute square of the Fourier transform of the LDOS. 

\subsection{Initial state from the ${\rm U}(1)$ basis: $|s\,m_z\rangle$ }
\label{initialU1}

\begin{figure}[htb]
\centering
\includegraphics*[width=3.5in]{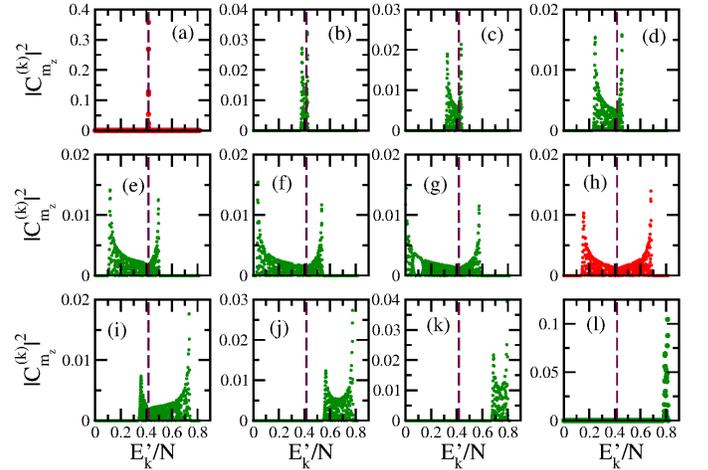}
\caption{(Color online) Structure of the ${\rm U}(1)$ basis vectors projected onto the eigenstates of the total Hamiltonian $H_s$; even parity, $N=2000$, $\xi=0.6$. The values of $m_z$ are: -1000 (a), -960  (b), -900 (c), -800 (d), -600 (e), -400 (f), -200 (g), 334 (h), 600 (i), 800 (j), 900 (k), 980 (l). Vertical dashed lines mark $E_{\text{ESQPT}}$. The states with energy closest to the separatrix are (a), with $e_{-1000}/N=0.4164$, and (h), with $e_{334}/N=0.4166$.}
\label{fig:basis}
\end{figure}

The dynamics can be anticipated by examining the structure of the initial states projected onto the energy eigenbasis, that is $|\Psi(0)\rangle = |s\,m_z\rangle = \sum_k C_{m_z}^{(k)} |\psi_k\rangle$. 
As expected from the previous analysis of the eigenstates, the ${\rm U}(1)$-ground state ($m_z=-N/2$) is highly localized in the eigenstate with $E'_k/N \sim E_{\text{ESQPT}}$, as seen in Fig.~\ref{fig:basis} (a). Its evolution should therefore be very slow, even though the energy $e'_{-N/2}/N$ of this state may be very high.  As $m_z$ increases from $-N/2$, the ${\rm U}(1)$-states become more and more delocalized in the energy eigenbasis [Figs. 9(b)-(h)], with higher contributions occurring at the edges of their energy distributions. The evolution should consequently become faster. Notice that due to this steady spreading in energy, the ${\rm U}(1)$ basis that has energy $e'_{m_z}/N$ closest to $E_{\text{ESQPT}}$, after the state with $m_z=-N/2$, is actually a very delocalized state with similar contributions from eigenstates below and above the separatrix. This is the state in Fig.~\ref{fig:basis} (h). Once the energies $e'_{m_z}/N$ get above the separatrix, the ${\rm U}(1)$-states gradually localize again, with higher contributions from eigenstates with large energies, that is those closer to the right edge of the spectrum [Figs.~\ref{fig:basis} (i), (j), (k), and (l)].

The symmetric shape of the ${\rm U}(1)$ basis vector in Fig.~\ref{fig:basis} (h) is similar to that found for the eigenstate in Fig.~2 (d), which is written in the ${\rm U}(1)$ basis and is above the separatrix, and for the eigenstate in Fig.~6 (a), which is written in the ${\rm SO}(2)$ basis and is below the separatrix. A closer look at the structures of these states reveal sinusoidal oscillations approximately modulated by a function $\propto (constant - E^2)^{-1/2}$. Interestingly, this envelope also coincides with the density of states of the XX model, as discussed below. This suggests a connection between the XX model and the LMG model, which is useful since several analytical results exist for the first one~\cite{Amico2008,Barouch1970}.

\subsubsection{Energy distribution of the initial state}

\begin{figure}[htb]
\centering
\includegraphics*[width=3.5in]{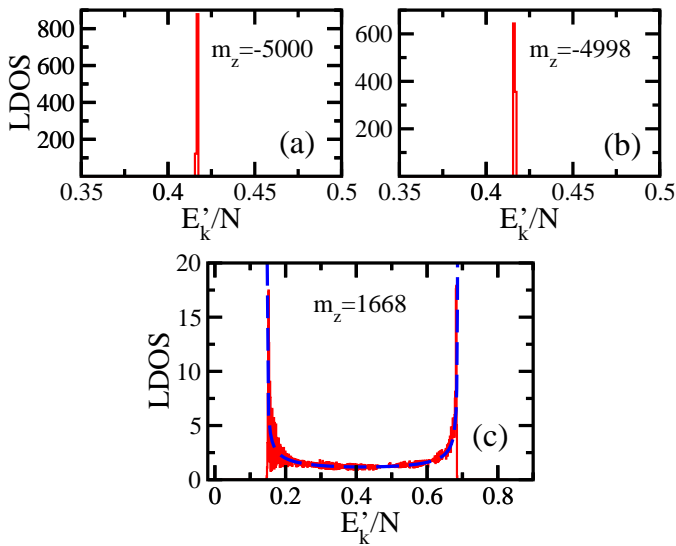}
\caption{(Color online) LDOS for initial states corresponding to ${\rm U}(1)$ basis vectors with $m_z=-N/2$ (a), $m_z=-N/2+2$ (b), and the one with the second closest $e'_{m_z}/N$ to $E_{\text{ESQPT}}$ (c); $N=10^4$, $\xi=0.6$. In (c), the dashed line represents Eq.(\ref{eq:fit}) with ${\cal A}^2 \sim 0.27$ and $e'_{m_z}/N = E_{\text{ESQPT}}$.}
\label{fig:LDOS}
\end{figure}

In what follows, we focus on the evolution of three initial states with even parity. They are the ones with $m_z=-N/2$ and  $m_z=-N/2+2$, and the one with the second closest $e'_{m_z}/N$ to $E_{\text{ESQPT}}$. 
The LDOS for these states are shown in Fig.~\ref{fig:LDOS}. For $m_z=-N/2$ and $m_z=-N/2+2$, the LDOS is highly localized on the eigenstates close to the separatrix. For the initial state with the second closest $e'_{m_z}/N$ to $E_{\text{ESQPT}}$, the LDOS  in Fig.~\ref{fig:LDOS} (c) is very similar to that found for the XX model with a single excitation. 

The Hamiltonian of the XX model is given by
\begin{equation}
H = \sum_{i} J \left( S_i^x S_{i+1}^x + S_i^y S_{i+1}^y \right),
\end{equation}
where $J$ is the coupling strength between nearest-neighbor spins. This is a noninteracting Hamiltonian that simply moves excitations along the chain. An excitation corresponds to a spin pointing up in the $z$ direction. For periodic boundary conditions and a single excitation, it is straightforward to obtain analytically the eigenvalues and eigenstates of this Hamiltonian, as shown in Appendix A. From these results, we can show that the LDOS for an initial state  $|\phi_l\rangle$ where all spins point down, except for one, is given by
\begin{equation}
\rho_{|\phi_l\rangle }(E) = \frac{1}{\pi \sqrt{J^2-E^2}}.
\label{eq:ldosXX}
\end{equation}
This expression, shifted by the energy $e'_{m_z}/N \sim E_{\text{ESQPT}}$, as
\begin{equation}
\rho_{\text{ini} }(E) = \frac{1}{\pi \sqrt{{\cal A}^2-(e'_{m_z}/N-E)^2}},
\label{eq:fit}
\end{equation}
fits the curve in Fig.~\ref{fig:LDOS} (c) extremely well. The only fitting parameter is ${\cal A}$, which is related with the range of energies sampled by the initial state.
It is interesting that the LDOS for a model with infinite-range interaction can coincide with that for a model with only nearest-neighbor couplings. 

\subsubsection{Survival Probability}

\begin{figure}[htb]
\centering
\includegraphics*[width=3.5in]{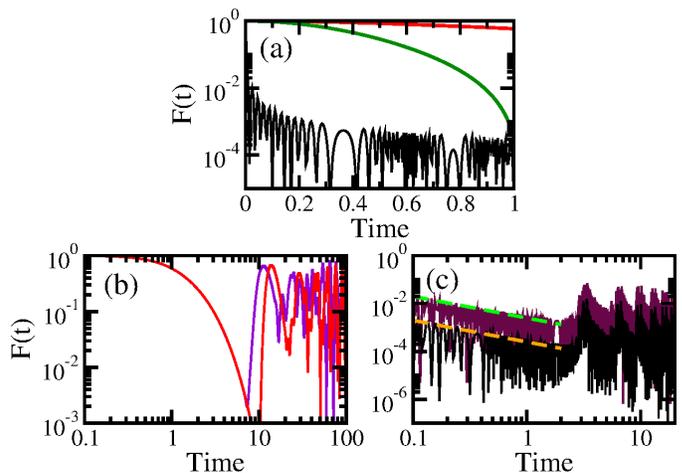}
\caption{(Color online) Survival Probability {\em vs} time. In (a) from top to bottom:  initial states corresponding to ${\rm U}(1)$ basis vectors with $m_z=-N/2$, $m_z=-N/2+2$, and the one with the second closest $e'_{m_z}/N$ to $E_{\text{ESQPT}}$; $N=10^4$. In (b): $m_z=-N/2$ for $N=10^3$ (first curve to show revival) and $N=10^4$. In (c): initial state with the second closest $e'_{m_z}/N$ to $E_{\text{ESQPT}}$; $N=10^3$ (top) and $N=10^4$ (bottom); dashed lines give $F(t) \propto 1/t$. All panels: $\xi=0.6$. Arbitrary units.}
\label{fig:Fid}
\end{figure}

The survival probability is shown in Fig.~\ref{fig:Fid} (a) for $N=10^4$. As predicted, the decay is very slow for $m_z=-N/2$ [top curve in Fig.~\ref{fig:Fid} (a)] and it becomes much faster as $m_z$ increases from $-N/2+2$ [middle curve] to $1668$ [bottom curve]. The latter is the state with the second closest $e'_{m_z}/N$ to $E_{\text{ESQPT}}$, for the parameters considered in the figure. 

Figure~\ref{fig:Fid} (b) reinforces the localization of the $m_z=-N/2$ state: the $F(t)$ curves for $N=10^3$ and $N=10^4$ fall on top of each other. In contrast, the short-time evolution of initial states with larger $m_z$ does accelerate (figure not shown).

In Fig.~\ref{fig:Fid} (c), we show the long-time evolution of the delocalized initial state with $e'_{m_z}/N \sim E_{\text{ESQPT}}$ for system sizes $N=10^3$ and $N=10^4$. The dashed lines represent a powerlaw decay $\propto t^{-1}$, which matches the numerical curves very well. This algebraic decay at long times can be justified by studying the Fourier transform of Eq.~(\ref{eq:fit}). It leads to the following expression for the survival probability,
\begin{eqnarray}
F(t) &=& \left| \frac{1}{\pi {\cal A}  }   
\int_{e'_{m_z}/N-{\cal A}}^{e'_{m_z}/N+{\cal A}}   
\frac{e^{-iEt} dE}{\sqrt{1 - \left(  \frac{E - e'_{m_z}/N}{ {\cal A} }  \right)^2} }
\right|^2 \\
&=& \left|  {\cal J}_0({\cal A} t)  \right|^2,
\end{eqnarray}
where ${\cal J}_0$ is the Bessel function of the first kind. For very long times,
\begin{equation}
F(t \gg {\cal A}) \simeq \frac{2}{\pi {\cal A} t} \cos^2 \left( {\cal A} t - \frac{\pi}{4} \right) ,
\end{equation}
from where the algebraic decay $\propto t^{-1}$ is evident. Beyond this decay, the survival probability fluctuates around a saturation value~\cite{Torres2014PRA,Torres2014NJP,Torres2014PRE,Torres2014PRAb,Torres2015,TavoraARXIV}.

The dynamics for the LMG model starting with a delocalized ${\rm U}(1)$-state with energy away from the ESQPT is therefore analogous to that impinged by the closed XX model on any initial state $|\phi_l\rangle $ with a single excitation. There are, however, evident differences between the two systems. (i) The speed of the evolution under the LMG Hamiltonian depends on the initial state, while for the XX case, it is the same for any  $|\phi_l\rangle $. (ii) For the LMG model, the density of states diverges at $E_{\text{ESQPT}}$,  while the shape of the level density for the XX model is equivalent to that of the LDOS in Eq.~(\ref{eq:ldosXX}), where divergences occur only at the edges of the spectrum~\cite{Iachello2015}. (iii) The Hamiltonian matrices for both models written in the basis of spins aligned in the $z$-direction are tridiagonal, but the structure of the LMG matrix is richer. From its analysis one can, in fact, identify the energy of the ESQPT critical point, as discussed below.

\subsubsection{Structure of the LMG Hamiltonian matrix}

\begin{figure}[htb]
\centering
\hspace{-0.5 cm}
\includegraphics*[width=3.3in]{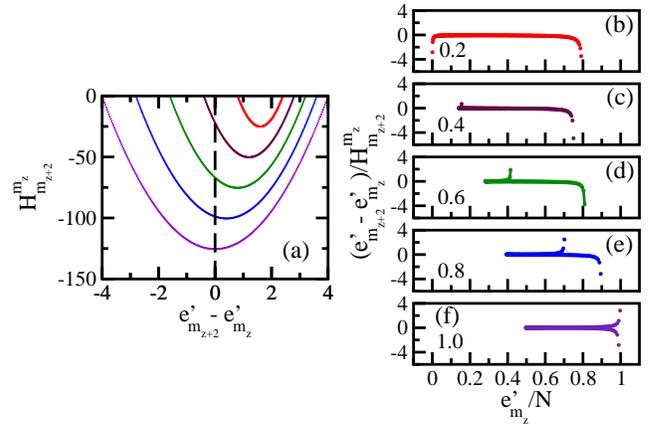}
\caption{(Color online) Structure of the Hamiltonian matrix of the LMG model, $H_{\text{s}}$ [Eq.~\ref{HtotalS}], written in the ${\rm U}(1)$ basis; only even parity is considered. In (a): coupling strength between two neighboring levels {\em vs} spacing between those levels; from top to bottom: $\xi=0.2, 0.4, 0.6, 0.8, 1.0$. In (b)-(f): ratio of the spacing between neighboring levels and their coupling strength; the value of $\xi$ is indicated in the panels. Absolute ratio $>1$ indicates ineffective coupling. Arbitrary units.}
\label{fig:structure}
\end{figure}

For the LMG model written in the ${\rm U}(1)$ basis with the diagonal elements ordered from the lowest to the highest value of $m_z$, the structure of the matrix for $\xi\leq \xi_c$ differs from that for $\xi > \xi_c$. This difference is explained in Fig.~\ref{fig:structure}.

Figure~\ref{fig:structure} (a) depicts the coupling strength between two neighboring levels, $H^{m_z}_{m_{z+2}} = \langle m_z |H_s | m_{z+2} \rangle $, {\em vs}  the spacing between the same two levels, $e'_{m_{z+2}} - e'_{m_{z}} = \langle m_{z+2} |H_s | m_{z+2} \rangle  - \langle m_z |H_s | m_{z} \rangle $.  For $\xi \leq \xi_c$, the spacing is always positive, which indicates that for any $m_z$, $e'_{m_{z+2}} > e'_{m_{z}} $ [see the top curve of Fig.~\ref{fig:structure} (a)]. The spacing is minimum at the edges of the matrix: for $m_z = - N/2$, where $e'_{m_{z}}$ has the lowest value, and for $m_z=N/2$, where  $e'_{m_{z}}$ has the highest value. In both cases, the magnitude of the coupling is close to zero and therefore ineffective. The ratio $\left( e'_{m_{z+2}} - e'_{m_{z}} \right)/\langle m_z |H_s | m_{z+2} \rangle$ between the level spacing and the coupling strength as a function of $e'_{m_{z}}$  is shown in Fig.~\ref{fig:structure} (b) for $\xi= \xi_c$. The absolute value of this ratio is indeed very large at the edges, so one expects the eigenstates to be highly localized at the borders  of the spectrum.

For $\xi > \xi_c$, the magnitude of the coupling strengths for the pairs of states ($m_z=-N/2$ and $m_z=-N/2+2$) and ($m_z=N/2$ and $m_z=N/2-2$) remain very close to zero and the absolute values of their spacings further increase [see Fig.~\ref{fig:structure} (a)]. Once again, one therefore expects the eigenstates with energies close to $e'_{m_{z}=-N/2}$ and $e'_{m_{z}=N/2}$ to be very localized. The difference with respect to the case where $\xi \leq \xi_c$ is that the spacings for $m_z$'s close to $-N/2$ have now negative values, indicating that $e'_{m_{z}=-N/2}$ is not the lowest energy anymore. The absolute value of the ratio $|(e'_{m_{z+2}} - e'_{m_{z}})/H^{m_z}_{m_{z+2}}|$ for $m_z=-N/2$ is still very large, but $e'_{m_{z}=-N/2}$ is now shifted to high values [see Figs.~\ref{fig:structure} (c)-(f)]. This value follows the separatrix, as discussed in Fig. 2(e). As a consequence, the energy of the localized eigenstate with $m_z \sim -N/2$ is now expected to also be away from the edge of the spectrum and to be $\sim E_{\text{ESQPT}}$. The presence of the ESQPT can therefore be anticipated even before diagonalization by performing this simple analysis of the matrix elements.

\subsubsection{Total magnetization in the $z$-direction}

\begin{figure}[htb]
\centering
\includegraphics*[width=3.5in]{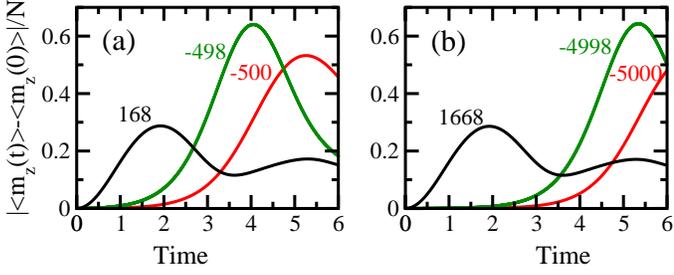}
\caption{(Color online) Evolution of the total magnetization in the $z$-direction. The values of $m_z(0)$ are indicated in the panels; they are $m_z(0)=-N/2$, $m_z(0)=-N/2+2$, and the one with the second closest $e'_{m_z}/N$ to $E_{\text{ESQPT}}$. In (a): $N=10^3$ and (b): $N=10^4$. Both panels: $\xi=0.6$. Arbitrary units.}
\label{fig:Obs}
\end{figure}

The different speeds of the evolution of ${\rm U}(1)$ basis vectors seen in Fig.~\ref{fig:Fid} must be reflected also in the dynamics of the total magnetization $m_z(t)$.
In Fig.~\ref{fig:Obs} , we show the evolution of the absolute value of the normalized difference $|m_z(t)-m_z(0) |/N$ starting with the same initial states considered in Fig.~\ref{fig:Fid}. The dynamics for the state with  $m_z=-N/2$ is, as expected, very slow and it further slows down as the system size increases from $N=10^3$ [Fig.~\ref{fig:Obs} (a)] to $N=10^4$ [Fig.~\ref{fig:Obs} (b)]. In contrast, the behavior of the state with the second closest energy to $E_{\text{ESQPT}}$, $m_z=168$ in (a) and $m_z=1668$ in (b), is very similar for different system sizes. The slow evolution of $m_z(t)$ signals the presence of the ESQPT.

\subsection{Initial state from the ${\rm SO}(2)$ basis: $|s\,m_x\rangle$ }
\label{initialSO2}

\begin{figure}[htb]
\centering
\hspace{-0.8 cm}
\includegraphics*[width=3.5in]{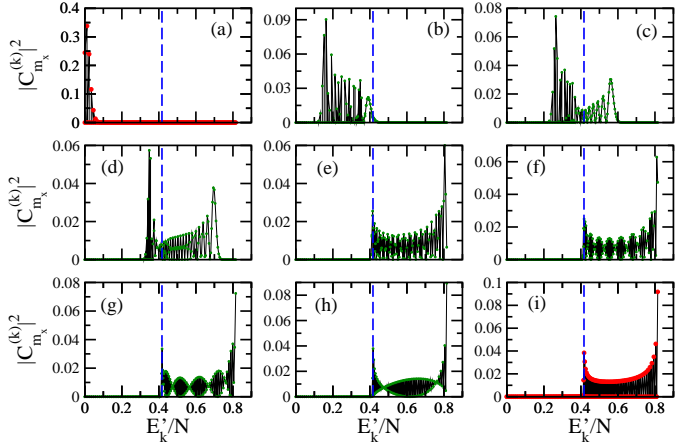}
\caption{(Color online)  Structure of the ${\rm SO}(2)$ basis vectors projected onto the eigenstates of the total Hamiltonian $H_s$;  $N=200$, $\xi=0.6$. The values of $m_x$ are: -100 (a), -75  (b), -58 (c), -39 (d), -10 (e), -5 (f), -3 (g), -1 (h), 0 (i). Circles are numerical results and thin black lines are guides for the eye. Vertical dashed lines mark $E_{\text{ESQPT}}$. Arbitrary units.}
\label{fig:CforSo2}
\end{figure}

We now consider as initial state, an eigenstate of the ${\rm SO}(2)$-part of the Hamiltonian, $|\Psi(0)\rangle = |s\,m_x\rangle = \sum_k C_{m_x}^{(k)} |\psi_k\rangle$. Equivalently to the analysis developed in Sec.~\ref{initialU1}, we start by studying in Fig.~\ref{fig:CforSo2} the dependence of the components $|C_{m_x}^{(k)} |^2$ on the eigenvalues of $H_s$. The structure is the same for states with a negative or positive value of $m_x$, so only negative values and $m_x=0$ are shown.

The state with $|m_x| =N/2$ is rather localized at the low eigenvalues of $H_s$. As $|m_x|$ increases, the states become more spread out and they move towards higher energies. Eventually, eigenstates with energies below and above the separatrix give significant contributions to $|s \, m_x \rangle$. The structures of the components below and above the separatrix are clearly different. As seen in Fig.~\ref{fig:CforSo2} (c), the damping of the oscillations above the separatrix is smoother and the frequency of the oscillations is smaller than below the separatrix. 

As $|m_x|$ approaches zero, the main contributions come from eigenstates with energies above $E_{\text{ESQPT}}$, where the structures of the eigenstates approach those of ${\rm U}(1)$ eigenstates, and very regular structures are formed [Figs.~\ref{fig:CforSo2} (f)-(h)]. At $m_x=0$  [Fig.~\ref{fig:CforSo2} (i)],  all contributing eigenstates have $E'_k/N>E_{\text{ESQPT}}$ and the nonzero values of $|C_{m_x}^{(k)} |^2$ have a dependence on energy very similar to that of the LDOS for the XX model given in Eq.~(\ref{eq:ldosXX}).

\begin{figure}[htb]
\centering
\includegraphics*[width=3.3in]{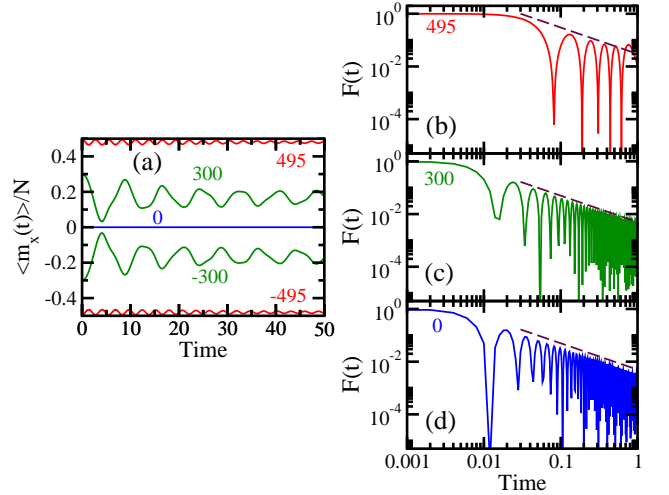}
\caption{(Color online)  Evolution of the total magnetization in the $x$-direction (left) and of the survival probability (right); $N=10^3$. The values of $m_x(0)$ are indicated in the panels. Dashed lines on the right panels correspond to  $F(t) \propto 1/t$. All panels: $\xi=0.6$. Arbitrary units.}
\label{fig:mx}
\end{figure}

The $x$-magnetization of the eigenstates that contribute to $|\Psi(0)\rangle = |s\,m_x\rangle$ have values close to the magnetization of the initial state.  As a result, the evolution of $m_x(t)$ is trapped around its initial value, as seen in Fig.~\ref{fig:mx} (a). The special case is that of $|\Psi(0)\rangle = |s\,0\rangle$, where only the eigenstates with $m_x=0$ lead to $|C_{m_x}^{(k)} |^2 \neq 0$.

The trapping of $m_x(t)$ is consistent with experimental studies of the phenomenon of bifurcation performed in Refs.~\cite{Zibold2010,Ferreira2013}. There, the initial state was a coherent state with a positive or negative value of $m_x(0)$. The behavior of $m_x(t)$ depended on the value of the control parameter. If the system was in the nonlinear regime, that is $\xi > \xi_c$, $m_x(t)$ remained trapped, oscillating around its initial value. If the system was in the linear regime, that is $\xi<\xi_c$, oscillations between both signs were verified and the temporal mean was zero. Here, we argue that distinct behaviors of $m_x(t)$ occur also for a fixed value of $\xi$, but for initial states prepared at different energies. If $|\Psi(0)\rangle$ is a superposition of energy eigenbasis with $E'_k > E_{\text{ESQPT}}$, then $m_x(t)=0$, since all contributing energy eigenbasis have $m_x=0$. In contrast, for a superposition of energy eigenbasis with $E'_k < E_{\text{ESQPT}}$,  the time average of $m_x(t)$ will be larger than zero (smaller than zero) if the majority of the contributions come from the branch of Fig.~\ref{fig:nt} (c) where the eigenstates have $m_x>0$ ($m_x<0$).

The right panels of Fig.~\ref{fig:mx} show the survival probability for the same initial states considered in Fig.~\ref{fig:mx} (a). The decay is slower for $m_x(0)=495$ [Fig.~\ref{fig:mx} (b)], because this state is more localized than the others, but apart from this, the decay is very similar for the three states. At long times, they show a powerlaw behavior $\propto t^{-1}$, as seen also for the delocalized state in Fig.~\ref{fig:Fid} (c). This was expected already from Fig.~\ref{fig:CforSo2}, which suggested that the LDOS for $|\Psi(0)\rangle = |s\,m_x\rangle$ with $|m_x| < N/2$, especially for those with $|m_x|$ very close to zero, should have a shape well described by Eq.~(\ref{eq:fit}).


\section{Concluding remarks}

Focusing on the LMG model, we identified several ways, other than the local divergence of the density of states, to detect the presence of an ESQPT. They are itemized below.

(i) The level of localization of the eigenstates written in the ${\rm U}(1)$ basis. At the separatrix, the eigenstates are highly localized in the ground state of the ${\rm U}(1)$-part of the LMG Hamiltonian.

(ii)  The ratio between the spacings of neighboring levels of the LMG Hamiltonian matrix written in the ${\rm U}(1)$ basis and their interaction strengths. One sees that these spacings  are larger than their coupling strengths for levels with energy very close to $E_{\text{ESQPT}}$. Since the coupling is ineffective, the eigenstate is localized. The ESQPT critical point can therefore be predicted even before diagonalization. 

(iii) The value of the total magnetization in the $z$-direction. The ground state of the ${\rm U}(1)$-part of the Hamiltonian has $m_z=-N/2$ ($n_t=0$). It is only for the localized eigenstates very close to the separatrix that the expectation value of the $z$-magnetization approaches this minimum value.

(iv) The bifurcation of the total magnetization in the $x$-direction.  The  structures of the eigenstates above the separatrix are closer to the ${\rm U}(1)$-symmetry and their $x$-magnetization is zero. Below the separatrix, the eigenstates are closer to the ${\rm SO}(2)$-symmetry. They come in pairs of degenerate states, each one having a positive or a negative value of $m_x$. The separatrix marks the point of this bifurcation.

(v) The speed of the evolution of ${\rm U}(1)$ basis vectors under the LMG Hamiltonian. The localization of the eigenstates at the separatrix implies that the evolution of the ${\rm U}(1)$ basis vector with $m_z=-N/2$  is very slow, as was confirmed by studying  $m_z(t)$ and the survival probability. This finding establishes a connection with experiments with ion traps~\cite{Jurcevic2014,Richerme2014}, where the evolution of ${\rm U}(1)$ basis vectors is currently studied. There, however, the range of the interaction is close to, but not exactly infinite. One of our future goals is to investigate whether the results obtained in this work can be extended to the scenario where $\alpha \neq 0$ in Hamiltonian (\ref{Hlmg}). 

The slow evolution despite the presence of infinite-range interactions emphasizes the importance of taking into account  both the Hamiltonian and also the initial state~\cite{SantosARXIV} when investigating nonequilibrium quantum dynamics. Conclusions based on only one of the two may result incomplete.

(vi) The trapping of the evolution of $m_x(t)$ close to its initial value. If the eigenstates contributing to the evolution of a chosen initial state have $E'_k/N <E_{\text{ESQPT}}$ and if they belong to a single branch of the two possible branches of values of $m_x$, the temporal mean of $m_x(t)$ will be nonzero. If the contributing eigenstates have $E'_k/N >E_{\text{ESQPT}}$, then the temporal mean of $m_x$ is zero. This analysis is similar to that developed in the experimental investigation of bifurcations with Bose-Einstein condensates~\cite{Zibold2010} and nuclear magnetic resonance~\cite{Ferreira2013}. The difference here is that the bifurcation occurs by varying the energy of the initial state, instead of by changing the value of the control parameter.

We also revealed similarities between the LDOS of the LMG model and the XX model with a single excitation. This allowed us to use the analytical expression obtained for the XX model to fit very well the LDOS of the LMG Hamiltonian. With it, we obtained an analytical expression for the long-time decay of the survival probability for both models, which is $\propto t^{-1}$.

It is our hope that the results reported in this work will motivate further experimental studies of ESQPTs, especially in the context of quench dynamics.


\begin{acknowledgments}
LFS and MT were supported by the NSF grant No.~DMR-1147430. FPB was funded by MINECO grant FIS2014-53448-C2-2-P and by Spanish Consolider-Ingenio 2010 (CPANCSD2007-00042). LFS and FPB thank Pedro P\'erez-Fern\'andez and Jorge Dukelsky for discussions, as well as the hospitality of Alejandro Frank and the Centro de Ciencias de la Complejidad (C$_3$)  at the UNAM in Mexico, where part of this work was carried out. We also
\end{acknowledgments}

\appendix
\section{XX Model}

The Hamiltonian of the XX model is given by
\begin{equation}
H = \sum_{i} J \left( S_i^x S_{i+1}^x + S_i^y S_{i+1}^y \right),
\end{equation}
where $J$ is the coupling strength between nearest-neighbor spins.  For periodic boundary conditions and a single excitation, the eigenvalues of this Hamiltonian can be found analytically as follows. Define the eigenstates as 
\begin{equation}
|\psi_k \rangle= \sum_{l=1}^N a_l^{(k)} |\phi_l\rangle,
\end{equation}
where  $|\phi_l\rangle$ is the state with a spin pointing up in the $z$-direction (an excitation) on site $l$, while all other spins point down. Substituting this equation and 
\begin{equation} 
H |\phi_l\rangle = \frac{J}{2} (|\phi_{l-1}\rangle  + |\phi_{l+1}\rangle )
\end{equation}
into $H|\psi_k \rangle = E_k |\psi_k \rangle$, gives the equation for the energy
\begin{equation}
E_k a_l^{(k)} = \frac{J}{2} (a_{l-1}^{(k)} + a_{l+1}^{(k)}).
\end{equation} 
Due to the periodic boundary conditions, $a_{l+N}^{(k)}=a_l^{(k)}$ and  it is appropriate to use the ansatz $a_l^{(k)} = e^{i 2 \pi k l/N}$, from where we obtain
\begin{equation}
E_k = J \cos \left(\frac{2 \pi k}{N}\right) ,
\label{eq:energy}
\end{equation} 
with $k=-N/2, -N/2+1, \ldots -1, 0,1,  \ldots N/2-1$, and the eigenstates,
\begin{equation}
|\psi_k \rangle = \frac{1}{\sqrt{N}}   \sum_{l=1}^N e^{i 2 \pi k l/N} |\phi_l\rangle ,
\end{equation}
which are Bloch waves.

For an initial state corresponding to one of the basis vectors $|\phi_l\rangle$, the LDOS is derived from $\rho_{|\phi_l\rangle }(E) = N^{-1} \sum_k \delta(E-E_k)$. In the thermodynamic limit, using $2 \pi k /N \rightarrow {\cal E}$, we have
\begin{equation}
\rho_{|\phi_l\rangle }(E) = \frac{1}{2\pi} \int_{-\pi}^{\pi} \delta(E-J \cos {\cal E}) d{\cal E} .
\end{equation}
The integral can be solved with the identity $\delta(f({\cal E})) = \sum_i \delta ({\cal E} - {\cal E}_i)/|f'({\cal E}_i)|$, where ${\cal E}_i = \pm \arccos(E/J)$ are the roots of $f({\cal E})$. We the obtain
\begin{equation}
\rho_{|\phi_l\rangle }(E) = \frac{1}{\pi \sqrt{J^2-E^2}}.
\label{eq:ldosXX_appendix}
\end{equation}
Notice that for the LDOS of the XX model, the probabilities $|a_l^{(k)}|^2=1$, while the components $|C_{m_z}^{(k)}|^2$ for the LMG model oscillate, as shown in Fig.~\ref{fig:basis} (h). Yet, the two resulting LDOS are comparable.



\begin{thebibliography}{74}
\expandafter\ifx\csname natexlab\endcsname\relax\def\natexlab#1{#1}\fi
\expandafter\ifx\csname bibnamefont\endcsname\relax
  \def\bibnamefont#1{#1}\fi
\expandafter\ifx\csname bibfnamefont\endcsname\relax
  \def\bibfnamefont#1{#1}\fi
\expandafter\ifx\csname citenamefont\endcsname\relax
  \def\citenamefont#1{#1}\fi
\expandafter\ifx\csname url\endcsname\relax
  \def\url#1{\texttt{#1}}\fi
\expandafter\ifx\csname urlprefix\endcsname\relax\def\urlprefix{URL }\fi
\providecommand{\bibinfo}[2]{#2}
\providecommand{\eprint}[2][]{\url{#2}}

\bibitem[{\citenamefont{Carr}(2011)}]{CarrBook}
\bibinfo{author}{\bibfnamefont{L.~D.} \bibnamefont{Carr}},
  \emph{\bibinfo{title}{Understanding Quantum Phase Transitions}}
  (\bibinfo{publisher}{CRC Press}, \bibinfo{address}{Boca Raton},
  \bibinfo{year}{2011}).

\bibitem[{\citenamefont{Sachdev}(2011)}]{SachdevBook}
\bibinfo{author}{\bibfnamefont{S.}~\bibnamefont{Sachdev}},
  \emph{\bibinfo{title}{Quantum Phase Transitions}}
  (\bibinfo{publisher}{Cambridge Press}, \bibinfo{address}{Cambridge},
  \bibinfo{year}{2011}).

\bibitem[{\citenamefont{Gilmore}(1979)}]{Gilmore1979}
\bibinfo{author}{\bibfnamefont{R.}~\bibnamefont{Gilmore}}, \bibinfo{journal}{J.
  Math. Phys.} \textbf{\bibinfo{volume}{20}}, \bibinfo{pages}{891}
  (\bibinfo{year}{1979}).

\bibitem[{\citenamefont{Feng et~al.}(1981)\citenamefont{Feng, Gilmore, and
  Deans}}]{Feng1981}
\bibinfo{author}{\bibfnamefont{D.~H.} \bibnamefont{Feng}},
  \bibinfo{author}{\bibfnamefont{R.}~\bibnamefont{Gilmore}}, \bibnamefont{and}
  \bibinfo{author}{\bibfnamefont{S.~R.} \bibnamefont{Deans}},
  \bibinfo{journal}{Phys. Rev. C} \textbf{\bibinfo{volume}{23}},
  \bibinfo{pages}{1254} (\bibinfo{year}{1981}).

\bibitem[{\citenamefont{Jaeger}(1998)}]{Jaeger1998}
\bibinfo{author}{\bibfnamefont{G.}~\bibnamefont{Jaeger}},
  \bibinfo{journal}{Arch. Hist. Exact Sci.} \textbf{\bibinfo{volume}{53}},
  \bibinfo{pages}{51 } (\bibinfo{year}{1998}).

\bibitem[{\citenamefont{Greiner et~al.}(2002)\citenamefont{Greiner, Mandel,
  Esslinger, H\"ansch, and Bloch}}]{Greiner2002}
\bibinfo{author}{\bibfnamefont{M.}~\bibnamefont{Greiner}},
  \bibinfo{author}{\bibfnamefont{O.}~\bibnamefont{Mandel}},
  \bibinfo{author}{\bibfnamefont{T.}~\bibnamefont{Esslinger}},
  \bibinfo{author}{\bibfnamefont{T.~W.} \bibnamefont{H\"ansch}},
  \bibnamefont{and} \bibinfo{author}{\bibfnamefont{I.}~\bibnamefont{Bloch}},
  \bibinfo{journal}{Nature} \textbf{\bibinfo{volume}{415}}, \bibinfo{pages}{39}
  (\bibinfo{year}{2002}).

\bibitem[{\citenamefont{Bloch et~al.}(2008)\citenamefont{Bloch, Dalibard, and
  Zwerger}}]{Bloch2008}
\bibinfo{author}{\bibfnamefont{I.}~\bibnamefont{Bloch}},
  \bibinfo{author}{\bibfnamefont{J.}~\bibnamefont{Dalibard}}, \bibnamefont{and}
  \bibinfo{author}{\bibfnamefont{W.}~\bibnamefont{Zwerger}},
  \bibinfo{journal}{Rev. Mod. Phys.} \textbf{\bibinfo{volume}{80}},
  \bibinfo{pages}{885} (\bibinfo{year}{2008}).

\bibitem[{\citenamefont{Baumann et~al.}(2010)\citenamefont{Baumann, Guerlin,
  Brennecke, and Esslinger}}]{Baumann2010}
\bibinfo{author}{\bibfnamefont{K.}~\bibnamefont{Baumann}},
  \bibinfo{author}{\bibfnamefont{C.}~\bibnamefont{Guerlin}},
  \bibinfo{author}{\bibfnamefont{F.}~\bibnamefont{Brennecke}},
  \bibnamefont{and}
  \bibinfo{author}{\bibfnamefont{T.}~\bibnamefont{Esslinger}},
  \bibinfo{journal}{Nature} \textbf{\bibinfo{volume}{464}},
  \bibinfo{pages}{1301} (\bibinfo{year}{2010}).

\bibitem[{\citenamefont{Cejnar et~al.}(2006)\citenamefont{Cejnar, Macek,
  Heinze, Jolie, and Dobe\~{s}}}]{Cejnar2006}
\bibinfo{author}{\bibfnamefont{P.}~\bibnamefont{Cejnar}},
  \bibinfo{author}{\bibfnamefont{M.}~\bibnamefont{Macek}},
  \bibinfo{author}{\bibfnamefont{S.}~\bibnamefont{Heinze}},
  \bibinfo{author}{\bibfnamefont{J.}~\bibnamefont{Jolie}}, \bibnamefont{and}
  \bibinfo{author}{\bibfnamefont{J.}~\bibnamefont{Dobe\~{s}}},
  \bibinfo{journal}{J. Phys. A} \textbf{\bibinfo{volume}{39}},
  \bibinfo{pages}{L515} (\bibinfo{year}{2006}).

\bibitem[{\citenamefont{Caprio et~al.}(2008)\citenamefont{Caprio, Cejnar, and
  Iachello}}]{Caprio2008}
\bibinfo{author}{\bibfnamefont{M.}~\bibnamefont{Caprio}},
  \bibinfo{author}{\bibfnamefont{P.}~\bibnamefont{Cejnar}}, \bibnamefont{and}
  \bibinfo{author}{\bibfnamefont{F.}~\bibnamefont{Iachello}},
  \bibinfo{journal}{Ann. of Phys.} \textbf{\bibinfo{volume}{323}},
  \bibinfo{pages}{1106 } (\bibinfo{year}{2008}).

\bibitem[{\citenamefont{P\'erez-Bernal and Iachello}(2008)}]{Bernal2008}
\bibinfo{author}{\bibfnamefont{F.}~\bibnamefont{P\'erez-Bernal}}
  \bibnamefont{and} \bibinfo{author}{\bibfnamefont{F.}~\bibnamefont{Iachello}},
  \bibinfo{journal}{Phys. Rev. A} \textbf{\bibinfo{volume}{77}},
  \bibinfo{pages}{032115} (\bibinfo{year}{2008}).

\bibitem[{\citenamefont{Cejnar and Jolie}(2009)}]{Cejnar2009}
\bibinfo{author}{\bibfnamefont{P.}~\bibnamefont{Cejnar}} \bibnamefont{and}
  \bibinfo{author}{\bibfnamefont{J.}~\bibnamefont{Jolie}},
  \bibinfo{journal}{Progr. Part. Nucl. Phys.} \textbf{\bibinfo{volume}{62}},
  \bibinfo{pages}{210 } (\bibinfo{year}{2009}).

\bibitem[{\citenamefont{P\'erez-Fern\'andez
  et~al.}(2011{\natexlab{a}})\citenamefont{P\'erez-Fern\'andez, Cejnar, Arias,
  Dukelsky, Garc\'{i}a-Ramos, and Rela\~no}}]{Fernandez2011}
\bibinfo{author}{\bibfnamefont{P.}~\bibnamefont{P\'erez-Fern\'andez}},
  \bibinfo{author}{\bibfnamefont{P.}~\bibnamefont{Cejnar}},
  \bibinfo{author}{\bibfnamefont{J.~M.} \bibnamefont{Arias}},
  \bibinfo{author}{\bibfnamefont{J.}~\bibnamefont{Dukelsky}},
  \bibinfo{author}{\bibfnamefont{J.~E.} \bibnamefont{Garc\'{i}a-Ramos}},
  \bibnamefont{and} \bibinfo{author}{\bibfnamefont{A.}~\bibnamefont{Rela\~no}},
  \bibinfo{journal}{Phys. Rev. A} \textbf{\bibinfo{volume}{83}},
  \bibinfo{pages}{033802} (\bibinfo{year}{2011}{\natexlab{a}}).

\bibitem[{\citenamefont{P\'erez-Fern\'andez
  et~al.}(2011{\natexlab{b}})\citenamefont{P\'erez-Fern\'andez, Rela\~no,
  Arias, Cejnar, Dukelsky, and Garc\'{i}a-Ramos}}]{Fernandez2011b}
\bibinfo{author}{\bibfnamefont{P.}~\bibnamefont{P\'erez-Fern\'andez}},
  \bibinfo{author}{\bibfnamefont{A.}~\bibnamefont{Rela\~no}},
  \bibinfo{author}{\bibfnamefont{J.~M.} \bibnamefont{Arias}},
  \bibinfo{author}{\bibfnamefont{P.}~\bibnamefont{Cejnar}},
  \bibinfo{author}{\bibfnamefont{J.}~\bibnamefont{Dukelsky}}, \bibnamefont{and}
  \bibinfo{author}{\bibfnamefont{J.~E.} \bibnamefont{Garc\'{i}a-Ramos}},
  \bibinfo{journal}{Phys. Rev. E} \textbf{\bibinfo{volume}{83}},
  \bibinfo{pages}{046208} (\bibinfo{year}{2011}{\natexlab{b}}).

\bibitem[{\citenamefont{Bastidas et~al.}(2014)\citenamefont{Bastidas,
  P\'erez-Fern\'andez, Vogl, and Brandes}}]{Bastidas2014}
\bibinfo{author}{\bibfnamefont{V.~M.} \bibnamefont{Bastidas}},
  \bibinfo{author}{\bibfnamefont{P.}~\bibnamefont{P\'erez-Fern\'andez}},
  \bibinfo{author}{\bibfnamefont{M.}~\bibnamefont{Vogl}}, \bibnamefont{and}
  \bibinfo{author}{\bibfnamefont{T.}~\bibnamefont{Brandes}},
  \bibinfo{journal}{Phys. Rev. Lett.} \textbf{\bibinfo{volume}{112}},
  \bibinfo{pages}{140408} (\bibinfo{year}{2014}).

\bibitem[{\citenamefont{P\'erez-Fern\'andez
  et~al.}(2009)\citenamefont{P\'erez-Fern\'andez, Rela\~no, Arias, Dukelsky,
  and Garc\'{i}a-Ramos}}]{Fernandez2009}
\bibinfo{author}{\bibfnamefont{P.}~\bibnamefont{P\'erez-Fern\'andez}},
  \bibinfo{author}{\bibfnamefont{A.}~\bibnamefont{Rela\~no}},
  \bibinfo{author}{\bibfnamefont{J.~M.} \bibnamefont{Arias}},
  \bibinfo{author}{\bibfnamefont{J.}~\bibnamefont{Dukelsky}}, \bibnamefont{and}
  \bibinfo{author}{\bibfnamefont{J.~E.} \bibnamefont{Garc\'{i}a-Ramos}},
  \bibinfo{journal}{Phys. Rev. A} \textbf{\bibinfo{volume}{80}},
  \bibinfo{pages}{032111} (\bibinfo{year}{2009}).

\bibitem[{\citenamefont{Yuan et~al.}(2012)\citenamefont{Yuan, Zhang, Li, Jing,
  and Kong}}]{Yuan2012}
\bibinfo{author}{\bibfnamefont{Z.-G.} \bibnamefont{Yuan}},
  \bibinfo{author}{\bibfnamefont{P.}~\bibnamefont{Zhang}},
  \bibinfo{author}{\bibfnamefont{S.-S.} \bibnamefont{Li}},
  \bibinfo{author}{\bibfnamefont{J.}~\bibnamefont{Jing}}, \bibnamefont{and}
  \bibinfo{author}{\bibfnamefont{L.-B.} \bibnamefont{Kong}},
  \bibinfo{journal}{Phys. Rev. A} \textbf{\bibinfo{volume}{85}},
  \bibinfo{pages}{044102} (\bibinfo{year}{2012}).

\bibitem[{\citenamefont{Brandes}(2013)}]{Brandes2013}
\bibinfo{author}{\bibfnamefont{T.}~\bibnamefont{Brandes}},
  \bibinfo{journal}{Phys. Rev. E} \textbf{\bibinfo{volume}{88}},
  \bibinfo{pages}{032133} (\bibinfo{year}{2013}).

\bibitem[{\citenamefont{Ribeiro et~al.}(2008)\citenamefont{Ribeiro, Vidal, and
  Mosseri}}]{Ribeiro2008}
\bibinfo{author}{\bibfnamefont{P.}~\bibnamefont{Ribeiro}},
  \bibinfo{author}{\bibfnamefont{J.}~\bibnamefont{Vidal}}, \bibnamefont{and}
  \bibinfo{author}{\bibfnamefont{R.}~\bibnamefont{Mosseri}},
  \bibinfo{journal}{Phys. Rev. E} \textbf{\bibinfo{volume}{78}},
  \bibinfo{pages}{021106} (\bibinfo{year}{2008}).

\bibitem[{\citenamefont{Bastarrachea-Magnani
  et~al.}(2014)\citenamefont{Bastarrachea-Magnani, Lerma-Hern\'andez, and
  Hirsch}}]{Bastarrachea2014b}
\bibinfo{author}{\bibfnamefont{M.~A.} \bibnamefont{Bastarrachea-Magnani}},
  \bibinfo{author}{\bibfnamefont{S.}~\bibnamefont{Lerma-Hern\'andez}},
  \bibnamefont{and} \bibinfo{author}{\bibfnamefont{J.~G.}
  \bibnamefont{Hirsch}}, \bibinfo{journal}{Phys. Rev. A}
  \textbf{\bibinfo{volume}{89}}, \bibinfo{pages}{032102}
  (\bibinfo{year}{2014}).

\bibitem[{\citenamefont{Ch\'avez-Carlos et~al.}()\citenamefont{Ch\'avez-Carlos,
  Bastarrachea-Magnani, Lerma-Hern\'andez, and Hirsch}}]{ChavezCarlosARXIV}
\bibinfo{author}{\bibfnamefont{J.}~\bibnamefont{Ch\'avez-Carlos}},
  \bibinfo{author}{\bibfnamefont{M.~A.} \bibnamefont{Bastarrachea-Magnani}},
  \bibinfo{author}{\bibfnamefont{S.}~\bibnamefont{Lerma-Hern\'andez}},
  \bibnamefont{and} \bibinfo{author}{\bibfnamefont{J.~G.}
  \bibnamefont{Hirsch}}, \bibinfo{note}{arXiv:1604.00725}.

\bibitem[{\citenamefont{Str\'ansk\'y et~al.}(2015)\citenamefont{Str\'ansk\'y,
  Macek, Leviatan, and Cejnar}}]{Stransky2015}
\bibinfo{author}{\bibfnamefont{P.}~\bibnamefont{Str\'ansk\'y}},
  \bibinfo{author}{\bibfnamefont{M.}~\bibnamefont{Macek}},
  \bibinfo{author}{\bibfnamefont{A.}~\bibnamefont{Leviatan}}, \bibnamefont{and}
  \bibinfo{author}{\bibfnamefont{P.}~\bibnamefont{Cejnar}},
  \bibinfo{journal}{Ann. of Phys.} \textbf{\bibinfo{volume}{356}},
  \bibinfo{pages}{57 } (\bibinfo{year}{2015}), ISSN \bibinfo{issn}{0003-4916}.

\bibitem[{\citenamefont{Winnewisser et~al.}(2005)\citenamefont{Winnewisser,
  Winnewisser, Medvedev, Behnke, De~Lucia, Ross, and Koput}}]{Winnewisser2005}
\bibinfo{author}{\bibfnamefont{B.~P.} \bibnamefont{Winnewisser}},
  \bibinfo{author}{\bibfnamefont{M.}~\bibnamefont{Winnewisser}},
  \bibinfo{author}{\bibfnamefont{I.~R.} \bibnamefont{Medvedev}},
  \bibinfo{author}{\bibfnamefont{M.}~\bibnamefont{Behnke}},
  \bibinfo{author}{\bibfnamefont{F.~C.} \bibnamefont{De~Lucia}},
  \bibinfo{author}{\bibfnamefont{S.~C.} \bibnamefont{Ross}}, \bibnamefont{and}
  \bibinfo{author}{\bibfnamefont{J.}~\bibnamefont{Koput}},
  \bibinfo{journal}{Phys. Rev. Lett.} \textbf{\bibinfo{volume}{95}},
  \bibinfo{pages}{243002} (\bibinfo{year}{2005}).

\bibitem[{\citenamefont{Zobov et~al.}(2006)\citenamefont{Zobov, Shirin,
  Polyansky, Tennyson, Coheur, Bernath, Carleer, and Colin}}]{Zobov2006}
\bibinfo{author}{\bibfnamefont{N.~F.} \bibnamefont{Zobov}},
  \bibinfo{author}{\bibfnamefont{S.~V.} \bibnamefont{Shirin}},
  \bibinfo{author}{\bibfnamefont{O.~L.} \bibnamefont{Polyansky}},
  \bibinfo{author}{\bibfnamefont{J.}~\bibnamefont{Tennyson}},
  \bibinfo{author}{\bibfnamefont{P.-F.} \bibnamefont{Coheur}},
  \bibinfo{author}{\bibfnamefont{P.~F.} \bibnamefont{Bernath}},
  \bibinfo{author}{\bibfnamefont{M.}~\bibnamefont{Carleer}}, \bibnamefont{and}
  \bibinfo{author}{\bibfnamefont{R.}~\bibnamefont{Colin}},
  \bibinfo{journal}{Chem. Phys. Lett.} \textbf{\bibinfo{volume}{414}},
  \bibinfo{pages}{193} (\bibinfo{year}{2006}).

\bibitem[{\citenamefont{Larese and Iachello}(2011)}]{Larese2011}
\bibinfo{author}{\bibfnamefont{D.}~\bibnamefont{Larese}} \bibnamefont{and}
  \bibinfo{author}{\bibfnamefont{F.}~\bibnamefont{Iachello}},
  \bibinfo{journal}{J. Mol. Struct.} \textbf{\bibinfo{volume}{1006}},
  \bibinfo{pages}{611 } (\bibinfo{year}{2011}).

\bibitem[{\citenamefont{Larese et~al.}(2013)\citenamefont{Larese,
  P\'erez-Bernal, and Iachello}}]{Larese2013}
\bibinfo{author}{\bibfnamefont{D.}~\bibnamefont{Larese}},
  \bibinfo{author}{\bibfnamefont{F.}~\bibnamefont{P\'erez-Bernal}},
  \bibnamefont{and} \bibinfo{author}{\bibfnamefont{F.}~\bibnamefont{Iachello}},
  \bibinfo{journal}{J. Mol. Struct.} \textbf{\bibinfo{volume}{1051}},
  \bibinfo{pages}{310 } (\bibinfo{year}{2013}).

\bibitem[{\citenamefont{Dietz et~al.}(2013)\citenamefont{Dietz, Iachello,
  Miski-Oglu, Pietralla, Richter, von Smekal, and Wambach}}]{Dietz2013}
\bibinfo{author}{\bibfnamefont{B.}~\bibnamefont{Dietz}},
  \bibinfo{author}{\bibfnamefont{F.}~\bibnamefont{Iachello}},
  \bibinfo{author}{\bibfnamefont{M.}~\bibnamefont{Miski-Oglu}},
  \bibinfo{author}{\bibfnamefont{N.}~\bibnamefont{Pietralla}},
  \bibinfo{author}{\bibfnamefont{A.}~\bibnamefont{Richter}},
  \bibinfo{author}{\bibfnamefont{L.}~\bibnamefont{von Smekal}},
  \bibnamefont{and} \bibinfo{author}{\bibfnamefont{J.}~\bibnamefont{Wambach}},
  \bibinfo{journal}{Phys. Rev. B} \textbf{\bibinfo{volume}{88}},
  \bibinfo{pages}{104101} (\bibinfo{year}{2013}).

\bibitem[{\citenamefont{Zhao et~al.}(2014)\citenamefont{Zhao, Jiang, Tang,
  Webb, and Liu}}]{Zhao2014}
\bibinfo{author}{\bibfnamefont{L.}~\bibnamefont{Zhao}},
  \bibinfo{author}{\bibfnamefont{J.}~\bibnamefont{Jiang}},
  \bibinfo{author}{\bibfnamefont{T.}~\bibnamefont{Tang}},
  \bibinfo{author}{\bibfnamefont{M.}~\bibnamefont{Webb}}, \bibnamefont{and}
  \bibinfo{author}{\bibfnamefont{Y.}~\bibnamefont{Liu}},
  \bibinfo{journal}{Phys. Rev. A} \textbf{\bibinfo{volume}{89}},
  \bibinfo{pages}{023608} (\bibinfo{year}{2014}).

\bibitem[{\citenamefont{Rela\~no et~al.}(2008)\citenamefont{Rela\~no, Arias,
  Dukelsky, Garc\'{i}a-Ramos, and P\'erez-Fern\'andez}}]{Relano2008}
\bibinfo{author}{\bibfnamefont{A.}~\bibnamefont{Rela\~no}},
  \bibinfo{author}{\bibfnamefont{J.~M.} \bibnamefont{Arias}},
  \bibinfo{author}{\bibfnamefont{J.}~\bibnamefont{Dukelsky}},
  \bibinfo{author}{\bibfnamefont{J.~E.} \bibnamefont{Garc\'{i}a-Ramos}},
  \bibnamefont{and}
  \bibinfo{author}{\bibfnamefont{P.}~\bibnamefont{P\'erez-Fern\'andez}},
  \bibinfo{journal}{Phys. Rev. A} \textbf{\bibinfo{volume}{78}},
  \bibinfo{pages}{060102} (\bibinfo{year}{2008}).

\bibitem[{\citenamefont{Engelhardt et~al.}(2015)\citenamefont{Engelhardt,
  Bastidas, Kopylov, and Brandes}}]{Engelhardt2015}
\bibinfo{author}{\bibfnamefont{G.}~\bibnamefont{Engelhardt}},
  \bibinfo{author}{\bibfnamefont{V.~M.} \bibnamefont{Bastidas}},
  \bibinfo{author}{\bibfnamefont{W.}~\bibnamefont{Kopylov}}, \bibnamefont{and}
  \bibinfo{author}{\bibfnamefont{T.}~\bibnamefont{Brandes}},
  \bibinfo{journal}{Phys. Rev. A} \textbf{\bibinfo{volume}{91}},
  \bibinfo{pages}{013631} (\bibinfo{year}{2015}).

\bibitem[{\citenamefont{Puebla and Rela\~no}(2015)}]{Puebla2015}
\bibinfo{author}{\bibfnamefont{R.}~\bibnamefont{Puebla}} \bibnamefont{and}
  \bibinfo{author}{\bibfnamefont{A.}~\bibnamefont{Rela\~no}},
  \bibinfo{journal}{Phys. Rev. E} \textbf{\bibinfo{volume}{92}},
  \bibinfo{pages}{012101} (\bibinfo{year}{2015}).

\bibitem[{\citenamefont{Santos and P\'erez-Bernal}(2015)}]{SantosBernal2015}
\bibinfo{author}{\bibfnamefont{L.~F.} \bibnamefont{Santos}} \bibnamefont{and}
  \bibinfo{author}{\bibfnamefont{F.}~\bibnamefont{P\'erez-Bernal}},
  \bibinfo{journal}{Phys. Rev. A} \textbf{\bibinfo{volume}{92}},
  \bibinfo{pages}{050101} (\bibinfo{year}{2015}).

\bibitem[{\citenamefont{P\'erez-Bernal and Santos}(2016)}]{Bernal2016}
\bibinfo{author}{\bibfnamefont{F.}~\bibnamefont{P\'erez-Bernal}}
  \bibnamefont{and} \bibinfo{author}{\bibfnamefont{L.~F.}
  \bibnamefont{Santos}}, \bibinfo{journal}{Fortschr. Phys.}
  (\bibinfo{year}{2016}), arXiv:1604.06851.

\bibitem[{\citenamefont{Iachello}(1981)}]{Iachello1981}
\bibinfo{author}{\bibfnamefont{F.}~\bibnamefont{Iachello}},
  \bibinfo{journal}{Chem.\ Phys.\ Lett.} \textbf{\bibinfo{volume}{78}},
  \bibinfo{pages}{581} (\bibinfo{year}{1981}).

\bibitem[{\citenamefont{Iachello and Levine}(1995)}]{IachelloBook}
\bibinfo{author}{\bibfnamefont{F.}~\bibnamefont{Iachello}} \bibnamefont{and}
  \bibinfo{author}{\bibfnamefont{R.~D.} \bibnamefont{Levine}},
  \emph{\bibinfo{title}{Algebraic Theory of Molecules}}
  (\bibinfo{publisher}{Oxford University Press}, \bibinfo{address}{Oxford},
  \bibinfo{year}{1995}).

\bibitem[{\citenamefont{Iachello and Oss.}(1996)}]{Iachello1996}
\bibinfo{author}{\bibfnamefont{F.}~\bibnamefont{Iachello}} \bibnamefont{and}
  \bibinfo{author}{\bibfnamefont{S.}~\bibnamefont{Oss.}}, \bibinfo{journal}{J.
  Chem. Phys.} \textbf{\bibinfo{volume}{104}}, \bibinfo{pages}{6956}
  (\bibinfo{year}{1996}).

\bibitem[{\citenamefont{P\'erez-Bernal
  et~al.}(2005)\citenamefont{P\'erez-Bernal, Santos, Vaccaro, and
  Iachello}}]{Bernal2005}
\bibinfo{author}{\bibfnamefont{F.}~\bibnamefont{P\'erez-Bernal}},
  \bibinfo{author}{\bibfnamefont{L.~F.} \bibnamefont{Santos}},
  \bibinfo{author}{\bibfnamefont{P.~H.} \bibnamefont{Vaccaro}},
  \bibnamefont{and} \bibinfo{author}{\bibfnamefont{F.}~\bibnamefont{Iachello}},
  \bibinfo{journal}{Chem. Phys. Lett.} \textbf{\bibinfo{volume}{414}},
  \bibinfo{pages}{398} (\bibinfo{year}{2005}).

\bibitem[{\citenamefont{Lipkin et~al.}(1965{\natexlab{a}})\citenamefont{Lipkin,
  Meshkov, and Glick}}]{Lipkin1965a}
\bibinfo{author}{\bibfnamefont{H.~J.} \bibnamefont{Lipkin}},
  \bibinfo{author}{\bibfnamefont{N.}~\bibnamefont{Meshkov}}, \bibnamefont{and}
  \bibinfo{author}{\bibfnamefont{A.~J.} \bibnamefont{Glick}},
  \bibinfo{journal}{Nucl. Phys.} \textbf{\bibinfo{volume}{62}},
  \bibinfo{pages}{188} (\bibinfo{year}{1965}{\natexlab{a}}).

\bibitem[{\citenamefont{Lipkin et~al.}(1965{\natexlab{b}})\citenamefont{Lipkin,
  Meshkov, and Glick}}]{Lipkin1965b}
\bibinfo{author}{\bibfnamefont{H.~J.} \bibnamefont{Lipkin}},
  \bibinfo{author}{\bibfnamefont{N.}~\bibnamefont{Meshkov}}, \bibnamefont{and}
  \bibinfo{author}{\bibfnamefont{A.~J.} \bibnamefont{Glick}},
  \bibinfo{journal}{Nucl. Phys.} \textbf{\bibinfo{volume}{62}},
  \bibinfo{pages}{199} (\bibinfo{year}{1965}{\natexlab{b}}).

\bibitem[{\citenamefont{Lipkin et~al.}(1965{\natexlab{c}})\citenamefont{Lipkin,
  Meshkov, and Glick}}]{Lipkin1965c}
\bibinfo{author}{\bibfnamefont{H.~J.} \bibnamefont{Lipkin}},
  \bibinfo{author}{\bibfnamefont{N.}~\bibnamefont{Meshkov}}, \bibnamefont{and}
  \bibinfo{author}{\bibfnamefont{A.~J.} \bibnamefont{Glick}},
  \bibinfo{journal}{Nucl. Phys.} \textbf{\bibinfo{volume}{62}},
  \bibinfo{pages}{211} (\bibinfo{year}{1965}{\natexlab{c}}).

\bibitem[{\citenamefont{Jurcevic et~al.}(2014)\citenamefont{Jurcevic, Lanyon,
  Hauke, Hempel, Zoller, Blatt, and Roos}}]{Jurcevic2014}
\bibinfo{author}{\bibfnamefont{P.}~\bibnamefont{Jurcevic}},
  \bibinfo{author}{\bibfnamefont{B.~P.} \bibnamefont{Lanyon}},
  \bibinfo{author}{\bibfnamefont{P.}~\bibnamefont{Hauke}},
  \bibinfo{author}{\bibfnamefont{C.}~\bibnamefont{Hempel}},
  \bibinfo{author}{\bibfnamefont{P.}~\bibnamefont{Zoller}},
  \bibinfo{author}{\bibfnamefont{R.}~\bibnamefont{Blatt}}, \bibnamefont{and}
  \bibinfo{author}{\bibfnamefont{C.~F.} \bibnamefont{Roos}},
  \bibinfo{journal}{Nature} \textbf{\bibinfo{volume}{511}},
  \bibinfo{pages}{202} (\bibinfo{year}{2014}).

\bibitem[{\citenamefont{Richerme et~al.}(2014)\citenamefont{Richerme, Gong,
  Lee, Senko, Smith, Foss-Feig, Michalakis, Gorshkov, and
  Monroe}}]{Richerme2014}
\bibinfo{author}{\bibfnamefont{P.}~\bibnamefont{Richerme}},
  \bibinfo{author}{\bibfnamefont{Z.-X.} \bibnamefont{Gong}},
  \bibinfo{author}{\bibfnamefont{A.}~\bibnamefont{Lee}},
  \bibinfo{author}{\bibfnamefont{C.}~\bibnamefont{Senko}},
  \bibinfo{author}{\bibfnamefont{J.}~\bibnamefont{Smith}},
  \bibinfo{author}{\bibfnamefont{M.}~\bibnamefont{Foss-Feig}},
  \bibinfo{author}{\bibfnamefont{S.}~\bibnamefont{Michalakis}},
  \bibinfo{author}{\bibfnamefont{A.~V.} \bibnamefont{Gorshkov}},
  \bibnamefont{and} \bibinfo{author}{\bibfnamefont{C.}~\bibnamefont{Monroe}},
  \bibinfo{journal}{Nature} \textbf{\bibinfo{volume}{511}},
  \bibinfo{pages}{198} (\bibinfo{year}{2014}).

\bibitem[{\citenamefont{Zibold et~al.}(2010)\citenamefont{Zibold, Nicklas,
  Gross, and Oberthaler}}]{Zibold2010}
\bibinfo{author}{\bibfnamefont{T.}~\bibnamefont{Zibold}},
  \bibinfo{author}{\bibfnamefont{E.}~\bibnamefont{Nicklas}},
  \bibinfo{author}{\bibfnamefont{C.}~\bibnamefont{Gross}}, \bibnamefont{and}
  \bibinfo{author}{\bibfnamefont{M.~K.} \bibnamefont{Oberthaler}},
  \bibinfo{journal}{Phys. Rev. Lett.} \textbf{\bibinfo{volume}{105}},
  \bibinfo{pages}{204101} (\bibinfo{year}{2010}).

\bibitem[{\citenamefont{Araujo-Ferreira
  et~al.}(2013)\citenamefont{Araujo-Ferreira, Auccaise, Sarthour, Oliveira,
  Bonagamba, and Roditi}}]{Ferreira2013}
\bibinfo{author}{\bibfnamefont{A.~G.} \bibnamefont{Araujo-Ferreira}},
  \bibinfo{author}{\bibfnamefont{R.}~\bibnamefont{Auccaise}},
  \bibinfo{author}{\bibfnamefont{R.~S.} \bibnamefont{Sarthour}},
  \bibinfo{author}{\bibfnamefont{I.~S.} \bibnamefont{Oliveira}},
  \bibinfo{author}{\bibfnamefont{T.~J.} \bibnamefont{Bonagamba}},
  \bibnamefont{and} \bibinfo{author}{\bibfnamefont{I.}~\bibnamefont{Roditi}},
  \bibinfo{journal}{Phys. Rev. A} \textbf{\bibinfo{volume}{87}},
  \bibinfo{pages}{053605} (\bibinfo{year}{2013}).

\bibitem[{\citenamefont{Trenkwalder}(2016)}]{Trenkwalder2016}
\bibinfo{author}{\bibfnamefont{A.}~\bibnamefont{Trenkwalder}},
  \bibinfo{journal}{Nat. Phys.}  (\bibinfo{year}{2016}), arXiv:1603.02979.

\bibitem[{\citenamefont{Amico et~al.}(2008)\citenamefont{Amico, Fazio,
  Osterloh, and Vedral}}]{Amico2008}
\bibinfo{author}{\bibfnamefont{L.}~\bibnamefont{Amico}},
  \bibinfo{author}{\bibfnamefont{R.}~\bibnamefont{Fazio}},
  \bibinfo{author}{\bibfnamefont{A.}~\bibnamefont{Osterloh}}, \bibnamefont{and}
  \bibinfo{author}{\bibfnamefont{V.}~\bibnamefont{Vedral}},
  \bibinfo{journal}{Rev. Mod. Phys.} \textbf{\bibinfo{volume}{80}},
  \bibinfo{pages}{517} (\bibinfo{year}{2008}).

\bibitem[{\citenamefont{Lerma and Dukelsky}(2013)}]{Lerma2013}
\bibinfo{author}{\bibfnamefont{S.~H.} \bibnamefont{Lerma}} \bibnamefont{and}
  \bibinfo{author}{\bibfnamefont{J.}~\bibnamefont{Dukelsky}},
  \bibinfo{journal}{Nuc. Phys. B} \textbf{\bibinfo{volume}{870}},
  \bibinfo{pages}{421 } (\bibinfo{year}{2013}).

\bibitem[{\citenamefont{Latorre et~al.}(2005)\citenamefont{Latorre, Or\'us,
  Rico, and Vidal}}]{Latorre2005}
\bibinfo{author}{\bibfnamefont{J.~I.} \bibnamefont{Latorre}},
  \bibinfo{author}{\bibfnamefont{R.}~\bibnamefont{Or\'us}},
  \bibinfo{author}{\bibfnamefont{E.}~\bibnamefont{Rico}}, \bibnamefont{and}
  \bibinfo{author}{\bibfnamefont{J.}~\bibnamefont{Vidal}},
  \bibinfo{journal}{Phys. Rev. A} \textbf{\bibinfo{volume}{71}},
  \bibinfo{pages}{064101} (\bibinfo{year}{2005}).

\bibitem[{\citenamefont{Barthel et~al.}(2006)\citenamefont{Barthel, Dusuel, and
  Vidal}}]{Barthel2006}
\bibinfo{author}{\bibfnamefont{T.}~\bibnamefont{Barthel}},
  \bibinfo{author}{\bibfnamefont{S.}~\bibnamefont{Dusuel}}, \bibnamefont{and}
  \bibinfo{author}{\bibfnamefont{J.}~\bibnamefont{Vidal}},
  \bibinfo{journal}{Phys. Rev. Lett.} \textbf{\bibinfo{volume}{97}},
  \bibinfo{pages}{220402} (\bibinfo{year}{2006}).

\bibitem[{\citenamefont{Frank and Isacker}(1994)}]{FrankBook}
\bibinfo{author}{\bibfnamefont{A.}~\bibnamefont{Frank}} \bibnamefont{and}
  \bibinfo{author}{\bibfnamefont{P.~V.} \bibnamefont{Isacker}},
  \emph{\bibinfo{title}{Algebraic Methods in Molecular and Nuclear Structure
  Physics}} (\bibinfo{publisher}{John Wiley and Sons, New York},
  \bibinfo{year}{1994}).

\bibitem[{\citenamefont{Iachello}(2015)}]{IachelloBook2}
\bibinfo{author}{\bibfnamefont{F.}~\bibnamefont{Iachello}},
  \emph{\bibinfo{title}{Lie Algebras and Applications}}
  (\bibinfo{publisher}{Springer}, \bibinfo{address}{Heidelberg},
  \bibinfo{year}{2015}).

\bibitem[{\citenamefont{Iachello and Arima}(1987)}]{IachelloBook3}
\bibinfo{author}{\bibfnamefont{F.}~\bibnamefont{Iachello}} \bibnamefont{and}
  \bibinfo{author}{\bibfnamefont{A.}~\bibnamefont{Arima}},
  \emph{\bibinfo{title}{The Interacting Boson Model}}
  (\bibinfo{publisher}{Cambridge Press}, \bibinfo{address}{Cambridge},
  \bibinfo{year}{1987}).

\bibitem[{\citenamefont{Stransky et~al.}(2014)\citenamefont{Stransky, Macek,
  and Cejnar}}]{Stransky2014}
\bibinfo{author}{\bibfnamefont{P.}~\bibnamefont{Stransky}},
  \bibinfo{author}{\bibfnamefont{M.}~\bibnamefont{Macek}}, \bibnamefont{and}
  \bibinfo{author}{\bibfnamefont{P.}~\bibnamefont{Cejnar}},
  \bibinfo{journal}{Ann. Phys.} \textbf{\bibinfo{volume}{345}},
  \bibinfo{pages}{73} (\bibinfo{year}{2014}).

\bibitem[{\citenamefont{Child}(1998)}]{Child1998}
\bibinfo{author}{\bibfnamefont{M.~S.} \bibnamefont{Child}},
  \bibinfo{journal}{J. Phys. A: Math. and Gen.} \textbf{\bibinfo{volume}{31}},
  \bibinfo{pages}{657} (\bibinfo{year}{1998}).

\bibitem[{\citenamefont{Zelevinsky et~al.}(1996)\citenamefont{Zelevinsky,
  Brown, Frazier, and Horoi}}]{ZelevinskyRep1996}
\bibinfo{author}{\bibfnamefont{V.}~\bibnamefont{Zelevinsky}},
  \bibinfo{author}{\bibfnamefont{B.~A.} \bibnamefont{Brown}},
  \bibinfo{author}{\bibfnamefont{N.}~\bibnamefont{Frazier}}, \bibnamefont{and}
  \bibinfo{author}{\bibfnamefont{M.}~\bibnamefont{Horoi}},
  \bibinfo{journal}{Phys. Rep.} \textbf{\bibinfo{volume}{276}},
  \bibinfo{pages}{85} (\bibinfo{year}{1996}).

\bibitem[{\citenamefont{Kota}(2001)}]{Kota2001}
\bibinfo{author}{\bibfnamefont{V.~K.~B.} \bibnamefont{Kota}},
  \bibinfo{journal}{Phys. Rep.} \textbf{\bibinfo{volume}{347}},
  \bibinfo{pages}{223} (\bibinfo{year}{2001}).

\bibitem[{\citenamefont{Santos et~al.}(2005)\citenamefont{Santos, Dykman,
  Shapiro, and Izrailev}}]{Santos2005loc}
\bibinfo{author}{\bibfnamefont{L.~F.} \bibnamefont{Santos}},
  \bibinfo{author}{\bibfnamefont{M.~I.} \bibnamefont{Dykman}},
  \bibinfo{author}{\bibfnamefont{M.}~\bibnamefont{Shapiro}}, \bibnamefont{and}
  \bibinfo{author}{\bibfnamefont{F.~M.} \bibnamefont{Izrailev}},
  \bibinfo{journal}{Phys. Rev. A} \textbf{\bibinfo{volume}{71}},
  \bibinfo{pages}{012317} (\bibinfo{year}{2005}).

\bibitem[{\citenamefont{Gubin and Santos}(2012)}]{Gubin2012}
\bibinfo{author}{\bibfnamefont{A.}~\bibnamefont{Gubin}} \bibnamefont{and}
  \bibinfo{author}{\bibfnamefont{L.~F.} \bibnamefont{Santos}},
  \bibinfo{journal}{Am. J. Phys.} \textbf{\bibinfo{volume}{80}},
  \bibinfo{pages}{246} (\bibinfo{year}{2012}).

\bibitem[{foo()}]{footPR}
\bibinfo{note}{We divided PR by $N$, but strictly, it should be divided by the
  dimension of the Hamiltonian matrix: $N/2+1$ for the even parity sector, when
  parity is taken into account. These differences are not important when $N$ is
  large.}

\bibitem[{\citenamefont{P\'erez-Bernal and \'Alvarez-Bajo}(2010)}]{Bernal2010}
\bibinfo{author}{\bibfnamefont{F.}~\bibnamefont{P\'erez-Bernal}}
  \bibnamefont{and}
  \bibinfo{author}{\bibfnamefont{O.}~\bibnamefont{\'Alvarez-Bajo}},
  \bibinfo{journal}{Phys. Rev. A} \textbf{\bibinfo{volume}{81}},
  \bibinfo{pages}{050101(R)} (\bibinfo{year}{2010}).

\bibitem[{\citenamefont{Shchesnovich and Konotop}(2009)}]{Shchesnovich2009}
\bibinfo{author}{\bibfnamefont{V.~S.} \bibnamefont{Shchesnovich}}
  \bibnamefont{and} \bibinfo{author}{\bibfnamefont{V.~V.}
  \bibnamefont{Konotop}}, \bibinfo{journal}{Phys. Rev. Lett.}
  \textbf{\bibinfo{volume}{102}}, \bibinfo{pages}{055702}
  (\bibinfo{year}{2009}).

\bibitem[{\citenamefont{Juli\'a-D\'{\i}az
  et~al.}(2010)\citenamefont{Juli\'a-D\'{\i}az, Dagnino, Lewenstein, Martorell,
  and Polls}}]{Diaz2010}
\bibinfo{author}{\bibfnamefont{B.}~\bibnamefont{Juli\'a-D\'{\i}az}},
  \bibinfo{author}{\bibfnamefont{D.}~\bibnamefont{Dagnino}},
  \bibinfo{author}{\bibfnamefont{M.}~\bibnamefont{Lewenstein}},
  \bibinfo{author}{\bibfnamefont{J.}~\bibnamefont{Martorell}},
  \bibnamefont{and} \bibinfo{author}{\bibfnamefont{A.}~\bibnamefont{Polls}},
  \bibinfo{journal}{Phys. Rev. A} \textbf{\bibinfo{volume}{81}},
  \bibinfo{pages}{023615} (\bibinfo{year}{2010}).

\bibitem[{\citenamefont{Trotzky et~al.}(2012)\citenamefont{Trotzky, Chen,
  Flesch, McCulloch, Schollw\"ock, Eisert, and Bloch}}]{Trotzky2012}
\bibinfo{author}{\bibfnamefont{S.}~\bibnamefont{Trotzky}},
  \bibinfo{author}{\bibfnamefont{Y.-A.} \bibnamefont{Chen}},
  \bibinfo{author}{\bibfnamefont{A.}~\bibnamefont{Flesch}},
  \bibinfo{author}{\bibfnamefont{I.~P.} \bibnamefont{McCulloch}},
  \bibinfo{author}{\bibfnamefont{U.}~\bibnamefont{Schollw\"ock}},
  \bibinfo{author}{\bibfnamefont{J.}~\bibnamefont{Eisert}}, \bibnamefont{and}
  \bibinfo{author}{\bibfnamefont{I.}~\bibnamefont{Bloch}},
  \bibinfo{journal}{Nature Phys.} \textbf{\bibinfo{volume}{8}},
  \bibinfo{pages}{325} (\bibinfo{year}{2012}).

\bibitem[{\citenamefont{Hazzard et~al.}(2014)\citenamefont{Hazzard, Gadway,
  Foss-Feig, Yan, Moses, Covey, Yao, Lukin, Ye, Jin et~al.}}]{Kaden2014}
\bibinfo{author}{\bibfnamefont{K.~R.~A.} \bibnamefont{Hazzard}},
  \bibinfo{author}{\bibfnamefont{B.}~\bibnamefont{Gadway}},
  \bibinfo{author}{\bibfnamefont{M.}~\bibnamefont{Foss-Feig}},
  \bibinfo{author}{\bibfnamefont{B.}~\bibnamefont{Yan}},
  \bibinfo{author}{\bibfnamefont{S.~A.} \bibnamefont{Moses}},
  \bibinfo{author}{\bibfnamefont{J.~P.} \bibnamefont{Covey}},
  \bibinfo{author}{\bibfnamefont{N.~Y.} \bibnamefont{Yao}},
  \bibinfo{author}{\bibfnamefont{M.~D.} \bibnamefont{Lukin}},
  \bibinfo{author}{\bibfnamefont{J.}~\bibnamefont{Ye}},
  \bibinfo{author}{\bibfnamefont{D.~S.} \bibnamefont{Jin}},
  \bibnamefont{et~al.}, \bibinfo{journal}{Phys. Rev. Lett.}
  \textbf{\bibinfo{volume}{113}}, \bibinfo{pages}{195302}
  (\bibinfo{year}{2014}).

\bibitem[{\citenamefont{Borgonovi et~al.}(2016)\citenamefont{Borgonovi,
  Izrailev, Santos, and Zelevinsky}}]{Borgonovi2016}
\bibinfo{author}{\bibfnamefont{F.}~\bibnamefont{Borgonovi}},
  \bibinfo{author}{\bibfnamefont{F.~M.} \bibnamefont{Izrailev}},
  \bibinfo{author}{\bibfnamefont{L.~F.} \bibnamefont{Santos}},
  \bibnamefont{and} \bibinfo{author}{\bibfnamefont{V.~G.}
  \bibnamefont{Zelevinsky}}, \bibinfo{journal}{Phys. Rep.}
  \textbf{\bibinfo{volume}{626}}, \bibinfo{pages}{1} (\bibinfo{year}{2016}).

\bibitem[{\citenamefont{Barouch et~al.}(1970)\citenamefont{Barouch, McCoy, and
  Dresden}}]{Barouch1970}
\bibinfo{author}{\bibfnamefont{E.}~\bibnamefont{Barouch}},
  \bibinfo{author}{\bibfnamefont{B.~M.} \bibnamefont{McCoy}}, \bibnamefont{and}
  \bibinfo{author}{\bibfnamefont{M.}~\bibnamefont{Dresden}},
  \bibinfo{journal}{Phys. Rev. A} \textbf{\bibinfo{volume}{2}},
  \bibinfo{pages}{1075} (\bibinfo{year}{1970}).

\bibitem[{\citenamefont{Torres-Herrera and
  Santos}(2014{\natexlab{a}})}]{Torres2014PRA}
\bibinfo{author}{\bibfnamefont{E.~J.} \bibnamefont{Torres-Herrera}}
  \bibnamefont{and} \bibinfo{author}{\bibfnamefont{L.~F.}
  \bibnamefont{Santos}}, \bibinfo{journal}{Phys. Rev. A}
  \textbf{\bibinfo{volume}{89}}, \bibinfo{pages}{043620}
  (\bibinfo{year}{2014}{\natexlab{a}}).

\bibitem[{\citenamefont{Torres-Herrera
  et~al.}(2014)\citenamefont{Torres-Herrera, Vyas, and Santos}}]{Torres2014NJP}
\bibinfo{author}{\bibfnamefont{E.~J.} \bibnamefont{Torres-Herrera}},
  \bibinfo{author}{\bibfnamefont{M.}~\bibnamefont{Vyas}}, \bibnamefont{and}
  \bibinfo{author}{\bibfnamefont{L.~F.} \bibnamefont{Santos}},
  \bibinfo{journal}{New J. Phys.} \textbf{\bibinfo{volume}{16}},
  \bibinfo{pages}{063010} (\bibinfo{year}{2014}).

\bibitem[{\citenamefont{Torres-Herrera and
  Santos}(2014{\natexlab{b}})}]{Torres2014PRE}
\bibinfo{author}{\bibfnamefont{E.~J.} \bibnamefont{Torres-Herrera}}
  \bibnamefont{and} \bibinfo{author}{\bibfnamefont{L.~F.}
  \bibnamefont{Santos}}, \bibinfo{journal}{Phys. Rev. E}
  \textbf{\bibinfo{volume}{89}}, \bibinfo{pages}{062110}
  (\bibinfo{year}{2014}{\natexlab{b}}).

\bibitem[{\citenamefont{Torres-Herrera and
  Santos}(2014{\natexlab{c}})}]{Torres2014PRAb}
\bibinfo{author}{\bibfnamefont{E.~J.} \bibnamefont{Torres-Herrera}}
  \bibnamefont{and} \bibinfo{author}{\bibfnamefont{L.~F.}
  \bibnamefont{Santos}}, \bibinfo{journal}{Phys. Rev. A}
  \textbf{\bibinfo{volume}{90}}, \bibinfo{pages}{033623}
  (\bibinfo{year}{2014}{\natexlab{c}}).

\bibitem[{\citenamefont{Torres-Herrera and Santos}(2015)}]{Torres2015}
\bibinfo{author}{\bibfnamefont{E.~J.} \bibnamefont{Torres-Herrera}}
  \bibnamefont{and} \bibinfo{author}{\bibfnamefont{L.~F.}
  \bibnamefont{Santos}}, \bibinfo{journal}{Phys. Rev. B}
  \textbf{\bibinfo{volume}{92}}, \bibinfo{pages}{014208}
  (\bibinfo{year}{2015}).

\bibitem[{\citenamefont{T\'avora et~al.}()\citenamefont{T\'avora,
  Torres-Herrera, and Santos}}]{TavoraARXIV}
\bibinfo{author}{\bibfnamefont{M.}~\bibnamefont{T\'avora}},
  \bibinfo{author}{\bibfnamefont{E.~J.} \bibnamefont{Torres-Herrera}},
  \bibnamefont{and} \bibinfo{author}{\bibfnamefont{L.~F.}
  \bibnamefont{Santos}}, \bibinfo{note}{arXiv:1601.05807}.

\bibitem[{\citenamefont{Iachello et~al.}(2015)\citenamefont{Iachello, Dietz,
  Miski-Oglu, and Richter}}]{Iachello2015}
\bibinfo{author}{\bibfnamefont{F.}~\bibnamefont{Iachello}},
  \bibinfo{author}{\bibfnamefont{B.}~\bibnamefont{Dietz}},
  \bibinfo{author}{\bibfnamefont{M.}~\bibnamefont{Miski-Oglu}},
  \bibnamefont{and} \bibinfo{author}{\bibfnamefont{A.}~\bibnamefont{Richter}},
  \bibinfo{journal}{Phys. Rev. B} \textbf{\bibinfo{volume}{91}},
  \bibinfo{pages}{214307} (\bibinfo{year}{2015}).

\bibitem[{\citenamefont{Santos et~al.}()\citenamefont{Santos, Borgonovi, and
  Celardo}}]{SantosARXIV}
\bibinfo{author}{\bibfnamefont{L.~F.} \bibnamefont{Santos}},
  \bibinfo{author}{\bibfnamefont{F.}~\bibnamefont{Borgonovi}},
  \bibnamefont{and} \bibinfo{author}{\bibfnamefont{G.~L.}
  \bibnamefont{Celardo}}, \bibinfo{note}{arXiv:1507.06649}.

\end{thebibliography}
\end{document}